%% file: SUS-14-007_temp.tex
\pdfoutput=1

\documentclass[11pt,twoside,a4paper,cmspaper,final,collab]{cms-tdr}
\def\svnVersion{345400}\def\cmsCernNoTag{CERN-EP/2016-008}\def\cmsCernDate{\today}\def\cmsMessage{Published in Physical Review D as \href{http://dx.doi.org/10.1103/PhysRevD.93.092009}{\doi{10.1103/PhysRevD.93.092009.}}}

\begin{document}\cmsNoteHeader{SUS-14-007}

\hyphenation{had-ron-i-za-tion}
\hyphenation{cal-or-i-me-ter}
\hyphenation{de-vices}
\RCS$Revision: 345280 $
\RCS$HeadURL: svn+ssh://svn.cern.ch/reps/tdr2/papers/SUS-14-007/trunk/SUS-14-007.tex $
\RCS$Id: SUS-14-007.tex 345280 2016-06-01 21:22:42Z nstrobbe $
\providecommand{\MR}{\ensuremath{M_R}\xspace}
\newlength\cmsFigWidth
\ifthenelse{\boolean{cms@external}}{\setlength\cmsFigWidth{0.85\columnwidth}}{\setlength\cmsFigWidth{0.4\textwidth}}
\ifthenelse{\boolean{cms@external}}{\providecommand{\cmsLeft}{top}}{\providecommand{\cmsLeft}{left}}
\ifthenelse{\boolean{cms@external}}{\providecommand{\cmsRight}{bottom}}{\providecommand{\cmsRight}{right}}
\ifthenelse{\boolean{cms@external}}{\providecommand{\cmsTable}[1]{\footnotesize #1}}{\providecommand{\cmsTable}[1]{\resizebox{\textwidth}{!}{#1}}}

\ifthenelse{\boolean{cms@external}}{\providecommand{\NA}{\ensuremath{\cdots}}\xspace}{\providecommand{\NA}{---\xspace}}
\ifthenelse{\boolean{cms@external}}{\providecommand{\CL}{C.L.\xspace}}{\providecommand{\CL}{CL\xspace}}

\cmsNoteHeader{SUS-14-007}
\title{\texorpdfstring{Search for supersymmetry in $\Pp \Pp$ collisions at $\sqrt{s}=8\TeV$ in final states with boosted $\PW$ bosons and $\cPqb$ jets using razor variables}{Search for supersymmetry in pp collisions at sqrt(s) = 8 TeV in final states with boosted W bosons and b jets using razor variables}}

\date{\today}

\abstract{
A search for supersymmetry in hadronic final states with highly boosted $\PW$ bosons and $\cPqb$ jets is presented, focusing on compressed scenarios.  The search is performed using proton-proton collision data at a center-of-mass energy of 8\TeV, collected by the CMS experiment at the LHC, corresponding to an integrated luminosity of 19.7\fbinv.  Events containing candidates for hadronic decays of boosted $\PW$ bosons are identified using jet substructure techniques, and are analyzed using the razor variables $\MR$ and $R^2$, which characterize a possible signal as a peak on a smoothly falling background.
The observed event yields in the signal regions are found to be consistent with the expected contributions from standard model processes, which are predicted using control samples in the data.  The results are interpreted in terms of gluino-pair production followed by their exclusive decay into top squarks and top quarks.  The analysis excludes gluino masses up to 1.1\TeV for light top squarks decaying solely to a charm quark and a neutralino, and up to 700\GeV for heavier top squarks decaying solely to a top quark and a neutralino.
}

\hypersetup{%
pdfauthor={CMS Collaboration},%
pdftitle={Search for supersymmetry in pp collisions at sqrt(s) = 8 TeV in final states with boosted W bosons and b jets using razor variables},%
pdfsubject={CMS},%
pdfkeywords={CMS, physics, SUSY, razor, boosted objects, b-tag, jet substructure}}

\maketitle

\section{Introduction}
\label{sec:introduction}

The CERN LHC has provided sufficient data to probe a large variety of theories beyond the standard model (SM).
Among these, theories based on supersymmetry (SUSY)~\cite{Wess,Golfand,Volkov,Chamseddine,Kane,Fayet,Barbieri,Hall,Ramond}, which predict the existence of a spectrum of supersymmetric partners to the SM particles, are strongly motivated.  Scenarios with nondegenerate supersymmetric particle spectra, with cross sections as low as ${\approx}1$~fb, have been explored in many final states; however, as yet no evidence for SUSY has been found.

The focus of many current searches is so-called natural SUSY~\cite{Barbieri:2009ev,Papucci:2011wy}, in which the Higgs boson mass can be stabilized without excessive fine-tuning.  In natural SUSY scenarios, the Higgsino mass parameter $\mu$ is required to be of the order of 100 GeV, and the lightest top squark $\PSQt_1$, the gluino $\PSg$, and the lightest bottom squark $\PSQb_1$ are constrained to have masses around the TeV scale, while the masses of the other superpartners are unconstrained and can be much heavier and beyond the LHC reach.  The possibility that the top squark could be light has motivated several searches by the ATLAS and CMS collaborations~\cite{Aad:2013ija,Aad:2014qaa,Aad:2014bva,Aad:2014kva,Aad:2014kra,Aad:2014mha,Aad:2014lra,Chatrchyan:2013xna,Chatrchyan:2013mya,Khachatryan:2014doa,Khachatryan:2015vra,Khachatryan:2015pot} for this sparticle.  In general, the sensitivity of these searches diminishes for direct top squark production when the mass of the top squark approaches that of the lightest supersymmetric particle (LSP), which is assumed to be the lightest neutralino $\PSGczDo$.   For searches that specifically target the decay $\PSQt_1 \to \cPqt \PSGczDo$, the sensitivity is reduced when the mass difference $\Delta m$ between the top squark and the LSP is comparable to the top quark mass $m_\cPqt$.

Here, we focus on two types of scenarios: the so-called compressed spectrum in which $\Delta m$ is very small, of the order of a few GeV to tens of GeV (\eg~\cite{Martin:2007gf, Martin:2007hn, Carena:2008mj}),
and  scenarios where $\Delta m \approx m_\cPqt$.  In the compressed case, the top squark decays to the LSP and soft decay products, which are difficult to detect.  When $\Delta m \approx m_\cPqt$, the signature of top squark production is very similar to that of $\cPqt\cPaqt$ production, which has a much higher cross section.
Therefore, to be sensitive to such processes, we cannot solely rely on the top squark decay products.
Possibilities to discriminate the signal are tagging the top squark events based on a jet from initial-state radiation (ISR) using the monojet signature~\cite{Khachatryan:2015wza,Aad:2014nra}, or searching for top squark events in cascade decays of heavier particles, such as the heavy top squark decays $\PSQt_2 \to \PSQt_1 + \PH/\cPZ$~\cite{Khachatryan:2014doa}, or from gluino decays.

In this paper, we search for the challenging top squark final states described above in gluino decays.
Specifically, we consider gluino-pair production where each gluino decays to a top squark and a top quark. We consider the scenarios in which the gluino has a mass of around 1\TeV and the lighter top squark has a mass of a few hundred\GeV.  Because of the significant mass gap between the gluino and the top squark, the top quark from the gluino decay will receive a large boost.  The top squark decays  to $\cPqc \PSGczDo$ for a small $\Delta m$, or to $\cPqt \PSGczDo$ for $\Delta m \approx m_\cPqt$, as
in the targeted searches for $\PSQt_1 \to \cPqt \PSGczDo$ mentioned above. The analysis described in
this paper is especially sensitive to the decay $\PSQt_1 \to \cPqc \PSGczDo$.  Consequently,
this analysis provides new information about the viability of natural SUSY.

\begin{figure*}[htbp]
\centering
\includegraphics[width=0.45\textwidth]{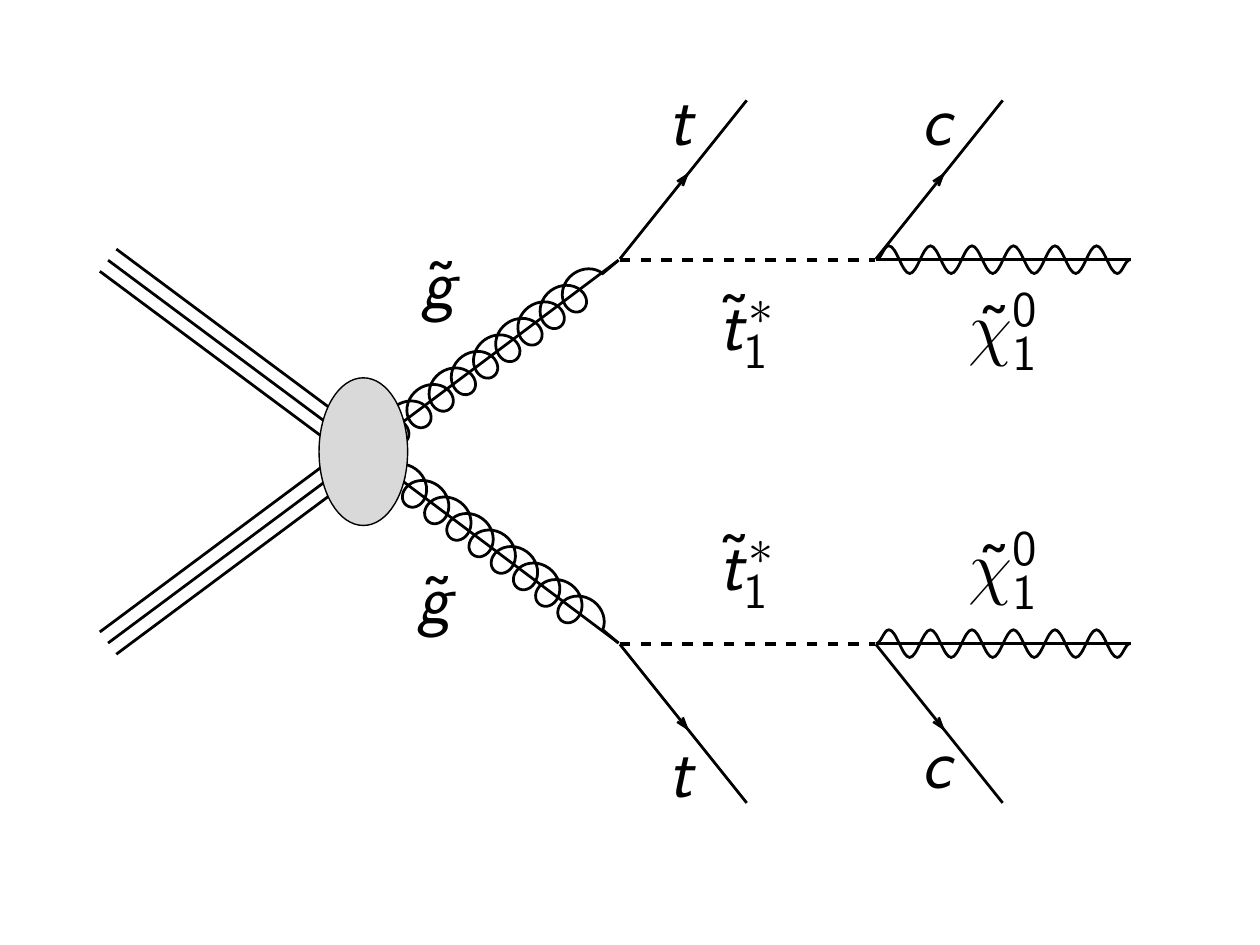}
\includegraphics[width=0.45\textwidth]{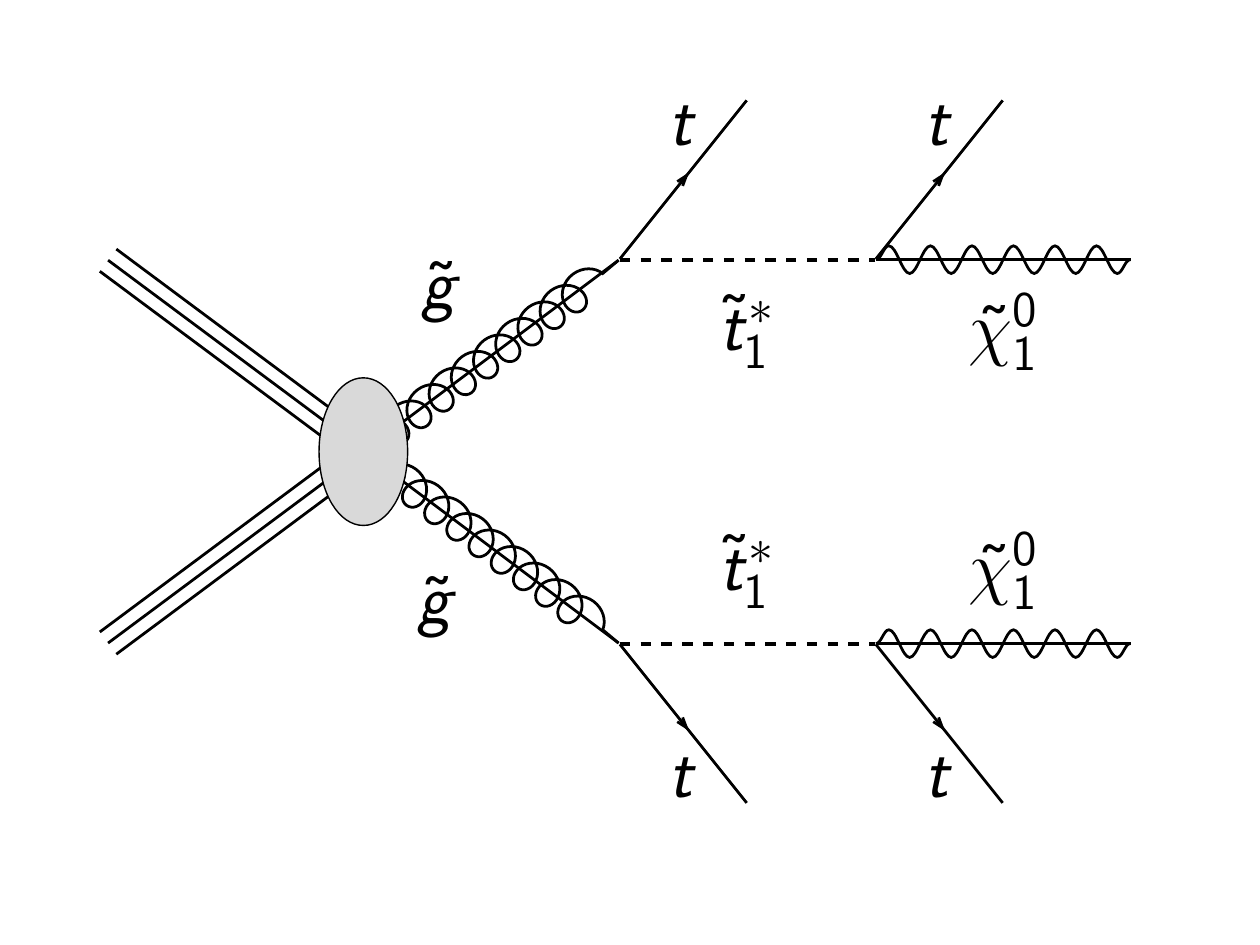}
\caption{Diagrams for the T1ttcc (left panel) and T1t1t (right panel) simplified model spectra. Here, an asterisk ($^*$) denotes an antiparticle of a supersymmetric partner.
\label{fig:diagrams}}
\end{figure*}

The gluino-pair production processes described above, with $\PSQt_1 \to \cPqc \PSGczDo$ or $\PSQt_1 \to \cPqt \PSGczDo$, can be described using simplified model spectra~\cite{Alves:2011wf,Alves:2011sq,Alwall:2008ag,Alwall:2008va,ArkaniHamed:2007fw,Chatrchyan:2013sza}. Specifically, the models T1ttcc and T1t1t, shown in Fig.~\ref{fig:diagrams}, are used in the design of the analysis and in the interpretation of the results.

In light of the discussion above, it is expected that boosted top quarks are a promising signature of new physics involving a massive gluino decaying to a relatively light top squark.
Boosted objects with high transverse momentum, $\pt$, are characterized by merged decay products separated by $ \Delta R \approx 2 m / \pt $, where $m$ denotes the mass of the decaying particle.
For the top quark decay products to be merged within the typical jet size of $\Delta R = 0.5$ requires a top quark momentum of ${\approx}700$\GeV, a value difficult to reach with proton-proton collisions at 8\TeV.
Therefore, in order to increase the signal efficiency by entering the boosted regime,
we focus on $\PW$ bosons from top quark decays, which require a more accessible $\pt$ of around $300$\GeV.
The targeted final state therefore contains boosted $\PW$ bosons and jets originating from $\cPqb$ quarks ($\cPqb$ jets) from top quark decays, light quark jets from unmerged hadronic $\PW$ boson decay products or charm quarks, and missing energy from the neutralinos.
Hadronically decaying boosted $\PW$ boson candidates are identified using the pruned jet mass~\cite{Ellis:2009su,Ellis:2009me,Chatrchyan:2013vbb} and a jet substructure observable
called N-subjettiness~\cite{Thaler:2010tr}.
The razor kinematic variables $\MR$ and $R^2$~\cite{rogan} are used to discriminate the processes with new heavy particles from SM processes in final states with jets and missing transverse energy.  To increase the sensitivity to new physics, we perform the analysis by partitioning the ($\MR$,$R^2$) plane into multiple bins.

This paper is organized as follows. The razor variables are introduced in Section~\ref{sec:razor}. Section~\ref{sec:cms} gives a brief overview of the CMS detector, while Section~\ref{sec:triggerdatasets} covers the triggers, data sets, and Monte Carlo (MC) simulated samples used in this analysis.
Details of the object definitions and event selection are given in Sections~\ref{sec:eventreco} and \ref{sec:selection}, respectively.
Section~\ref{sec:Wtag_SF} describes the data/simulation scale factors that are needed to correct the modeling of the boosted $\PW$ boson tagger.
The statistical analysis is explained in Section~\ref{sec:likelihood}, and Section~\ref{sec:systematics} covers the systematic uncertainties.  Finally, our results and their interpretation are presented in Section~\ref{sec:interpretation}, followed by a summary in Section~\ref{sec:summary}.

\section{Razor variables \label{sec:razor}}

The razor variables $\MR$ and $R^2$ \cite{rogan} are useful for
describing a signal arising from the pair production of heavy particles, each of which
decays to a massless visible particle and a massive invisible particle.
In the two-dimensional razor plane, a signal with heavy particles is expected to appear as a peak on top of smoothly falling SM backgrounds, which can be empirically described using exponential functions.
For this reason, the razor variables are robust discriminators for SUSY signals in which supersymmetric particles are pair produced and decay to SM particles and the LSP.  For the simple case in which the final state
comprises two visible particles, \eg jets, the razor variables are defined using the momenta $\vec{p}^{\,\mathrm{j}_1}$ and $\vec{p}^{\,\mathrm{j}_2}$ of the two jets as
\begin{eqnarray}
\label{eq:MRstar}
 \MR & \equiv &
 \sqrt{
   (|\vec{p}^{\,\mathrm{j}_1}|+|\vec{p}^{\,\mathrm{j}_2}|)^2 -
   (p^{\mathrm{j}_1}_z+p^{\mathrm{j}_2}_z)^2}
 \; ,\\
 M_\mathrm{T}^\mathrm{R} & \equiv & \sqrt{ \frac{\ETm(\pt^{\mathrm{j}_1}+\pt^{\mathrm{j}_2}) -
      \ptvecmiss {\cdot}
      (\ptvec^{\,\mathrm{j}_1}+\ptvec^{\,\mathrm{j}_2})}{2}} \; ,
\label{eq:MRT}
\end{eqnarray}
where $p^{\mathrm{j}_{1,2}}_z$ are the $z$ components of the $\mathrm{j}_{1,2}$ momenta, \ptvecmiss is the missing transverse momentum, computed as the negative vector sum of the transverse momenta of all observed particles in the event, and \ETm is its magnitude (see Section~\ref{sec:eventreco} for a more precise definition).
Given $\MR$ and the transverse quantity $M^\mathrm{R}_\mathrm{T}$, the razor dimensionless ratio is defined as
\begin{eqnarray}
R \equiv \frac{M_\mathrm{T}^\mathrm{R}}{\MR}
\; .
\label{eq:R2}
\end{eqnarray}
If the heavy mother particle is denoted by $G$ and the heavy invisible daughter particle is denoted by $\chi$, the peak of the $\MR$ distribution and end point of the $M_\mathrm{T}^\mathrm{R}$ distribution are both estimates of the quantity $(m_\mathrm{G}^\mathrm{2} - m_{\chi}^\mathrm{2})/m_\mathrm{G}$.
When the decay chains are complicated, producing multiple particles in the final state,
the razor variables can still be meaningfully calculated by reducing the final state to a two-``megajet'' structure.
The megajet algorithm aims to cluster visible particles coming from the decays of the same heavy supersymmetric particle.  The razor variables $\MR$ and $R^2$ are computed using the four-momenta of the two megajets, where the megajet four-momentum is the sum of the four-momenta of the particles comprising the megajet.  Studies show that, of all the possible clusterings, the one that minimizes the sum of the squared invariant masses of the megajets maximizes the efficiency with which particles are matched to their heavy supersymmetric particle ancestor~\cite{razorPRL}.

\begin{figure*}[tpb]
\centering
\includegraphics[width=0.49\textwidth]{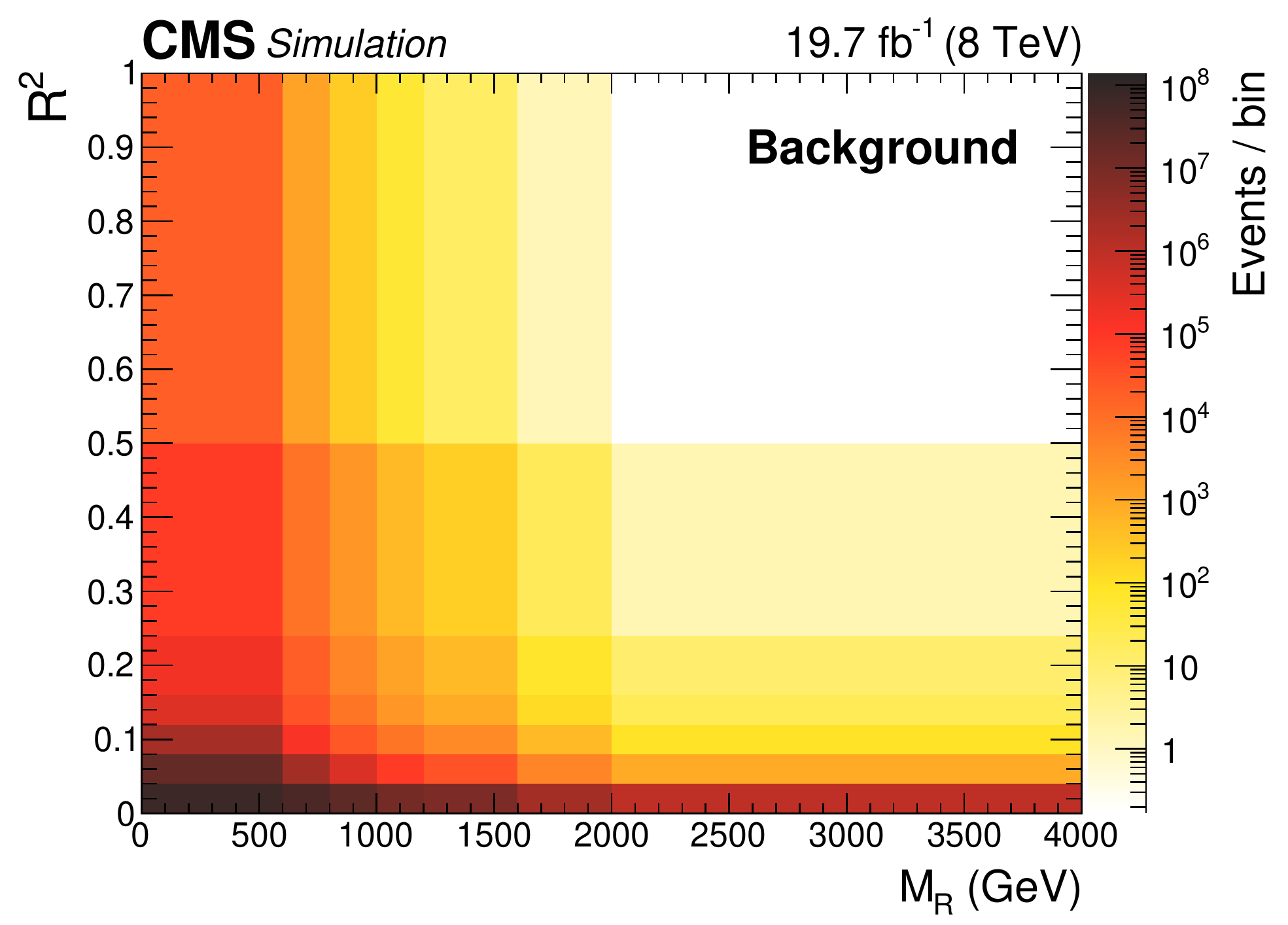}
\includegraphics[width=0.49\textwidth]{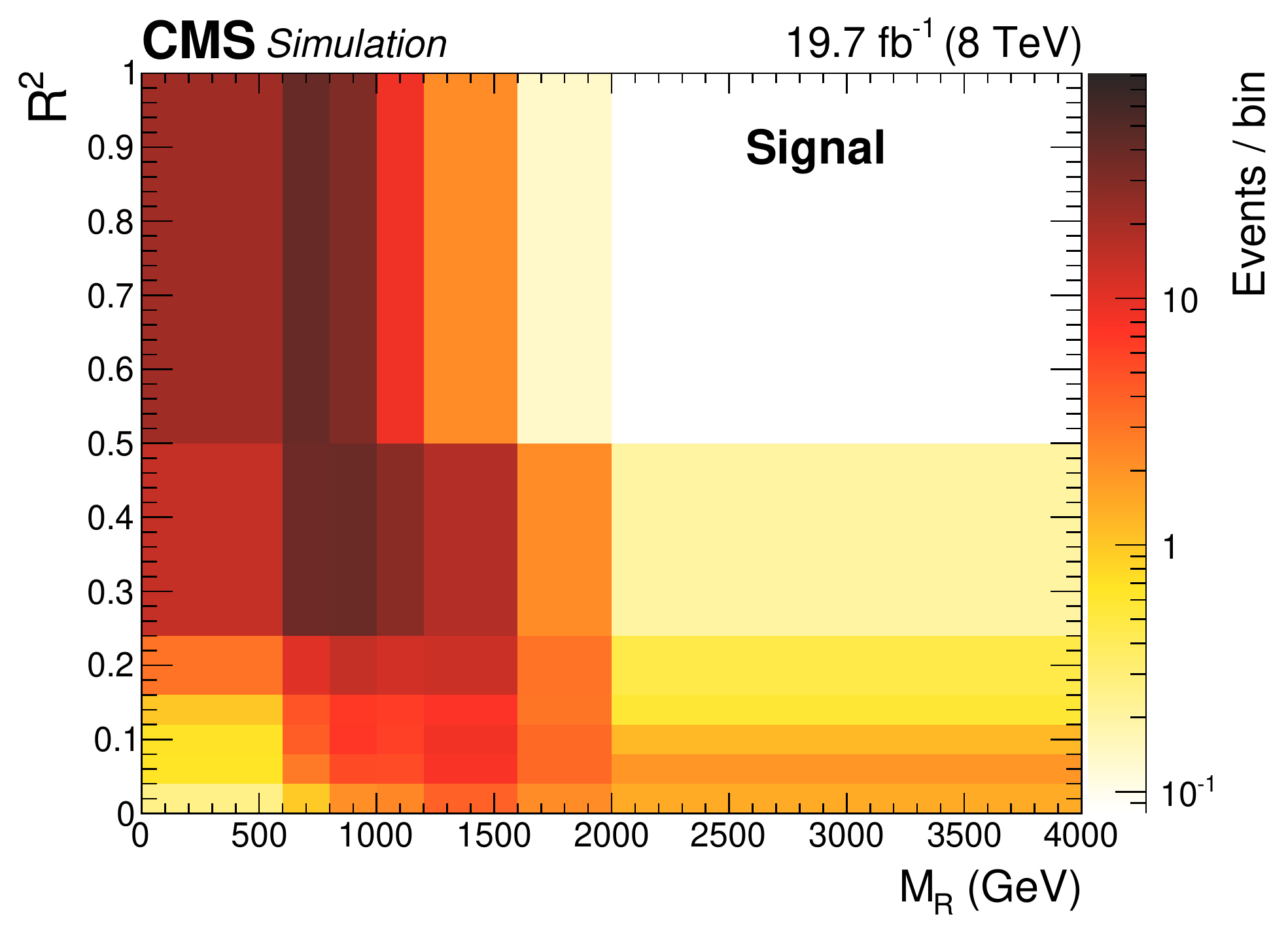}
\caption{Distributions in the ($\MR$,$R^2$) space of the overall SM backgrounds and a T1ttcc signal with $m_{\PSg} =1\TeV$, $m_{\PSQt} =325\GeV$ and $m_{\PSGczDo} =300\GeV$, both obtained from simulation.  A very loose selection is used:  a good primary vertex and at least three jets, one of which is required to have $\pt > 200$\GeV.
\label{fig:MRR2baseline}}
\end{figure*}

Figure~\ref{fig:MRR2baseline} shows the simulated distributions of the overall SM background and a T1ttcc signal with $m_{\PSg} = 1\TeV$, $m_{\PSQt} = 325\GeV$, and $m_{\PSGczDo} = 300\GeV$ in the ($\MR$,$R^2$) plane.  The binning is chosen in accordance with the exponentially falling behavior of the razor variables, to optimize the statistical precision in each bin.  The numerical values for the bin boundaries, which are used all through the analysis are given in Table~\ref{tab:results_prediction}.
The SM background, which mainly arises from multijet production, is dominant at low values of $R^2$, while the SUSY-like signal peaks higher in the ($\MR$,$R^2$) plane ($\MR$ peaks at around 900\GeV, which is the expected value).

In order to be sensitive to low-\ETm scenarios (small $\Delta m$), we use a lower $R^2$ threshold than that used in previous razor analyses~\cite{razor2010,razorPRL,Chatrchyan:2014goa,Khachatryan:2015pwa}.
To exploit the boosted phase space in which the expected signal significance is greater than in the nonboosted phase space, we work at large $(m_\mathrm{G}^\mathrm{2} - m_{\chi}^\mathrm{2})/m_\mathrm{G}$ and thus at high $\MR$, allowing us to raise the $\MR$ threshold.
This has the added virtue of keeping the SM backgrounds at a manageable level.

\section{The CMS detector \label{sec:cms}}

A detailed description of the CMS detector, together with a definition of the coordinate system used and the relevant kinematic variables, can be found elsewhere~\cite{Chatrchyan:2008zzk}.  A characteristic feature of the CMS detector is its superconducting solenoid magnet, of 6\unit{m} internal diameter, which provides a field of 3.8\unit{T}.  Within the field volume are a silicon pixel and strip tracker, a lead tungstate crystal electromagnetic calorimeter, and a brass and scintillator hadron calorimeter.  Muon detectors based on gas-ionization chambers are embedded in a steel flux-return yoke located outside the solenoid.
Events are collected by a two-layer trigger system, where the first level is composed of custom hardware processors,
and is followed by a software-based high-level trigger.

The tracking system covers the pseudorapidity region $\abs{\eta} < 2.5$, the muon detector $\abs{\eta} < 2.4$, and the calorimeters $\abs{\eta} < 3.0$.  Additionally, the forward region at $3 < \abs{\eta} < 5$ is covered by steel and quartz fiber forward calorimeters.  The near hermeticity of the detector permits an accurate measurement of the momentum balance in the transverse plane.

\section{Trigger and event samples \label{sec:triggerdatasets}}

This analysis is based on a sample of proton-proton collision data at $\sqrt{s}=8\TeV$ collected by the
CMS experiment in 2012 and corresponding to an integrated luminosity of 19.7\fbinv.
Events are selected using two triggers,
requiring either the highest jet \pt or the scalar sum $\HT$ of jet transverse momenta to be above given thresholds.  The jet \pt threshold was 320\GeV (and 400\GeV for a brief data taking period corresponding to 1.8\fbinv), while the $\HT$ threshold was 650\GeV.
The two trigger algorithms were based on a fast implementation of the particle-flow (PF) reconstruction method~\cite{PF2,PF}, which is described in Section~\ref{sec:eventreco}.

\begin{figure*}[htpb]
\centering
\includegraphics[width=0.49\textwidth]{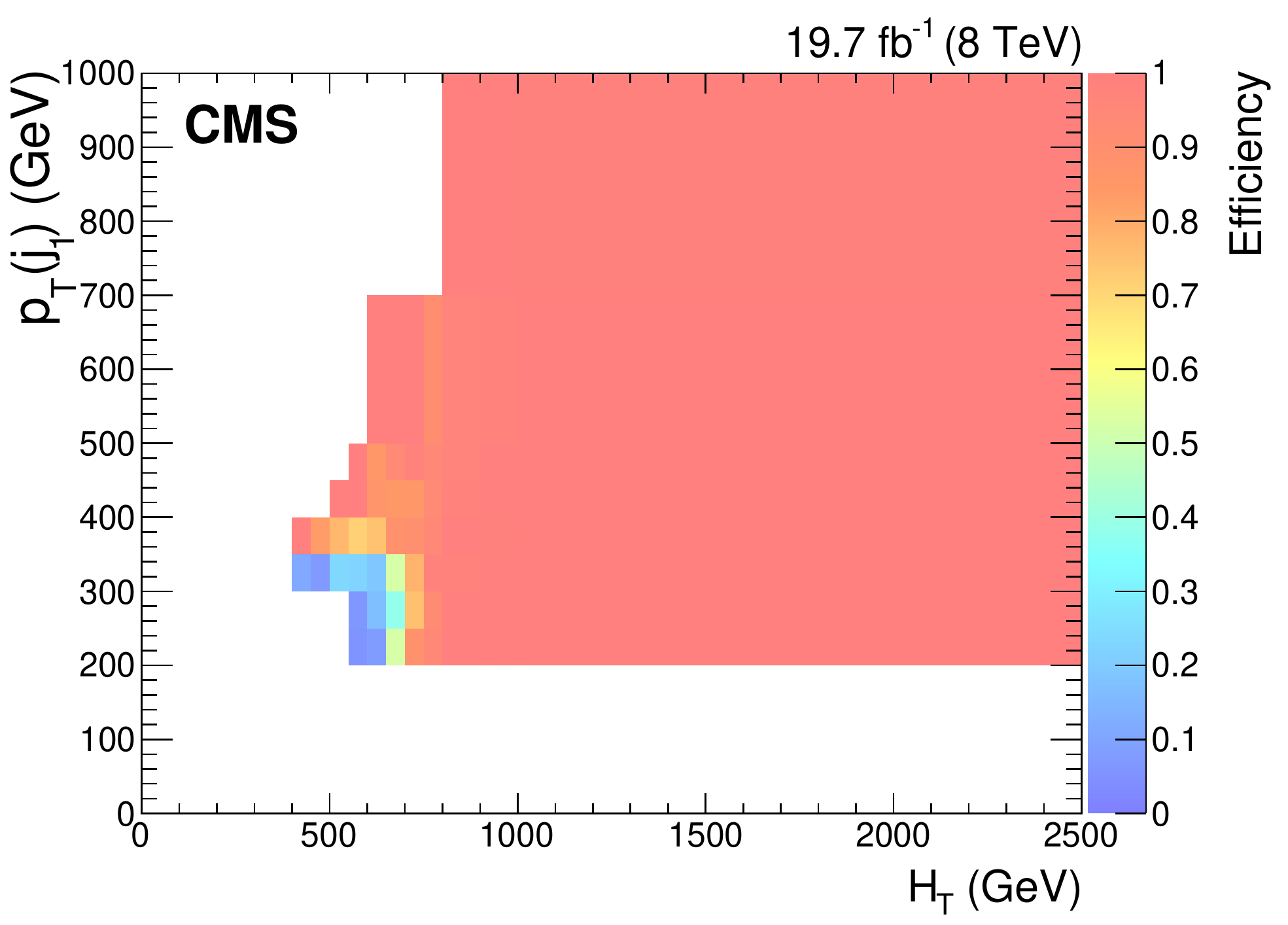}
\includegraphics[width=0.49\textwidth]{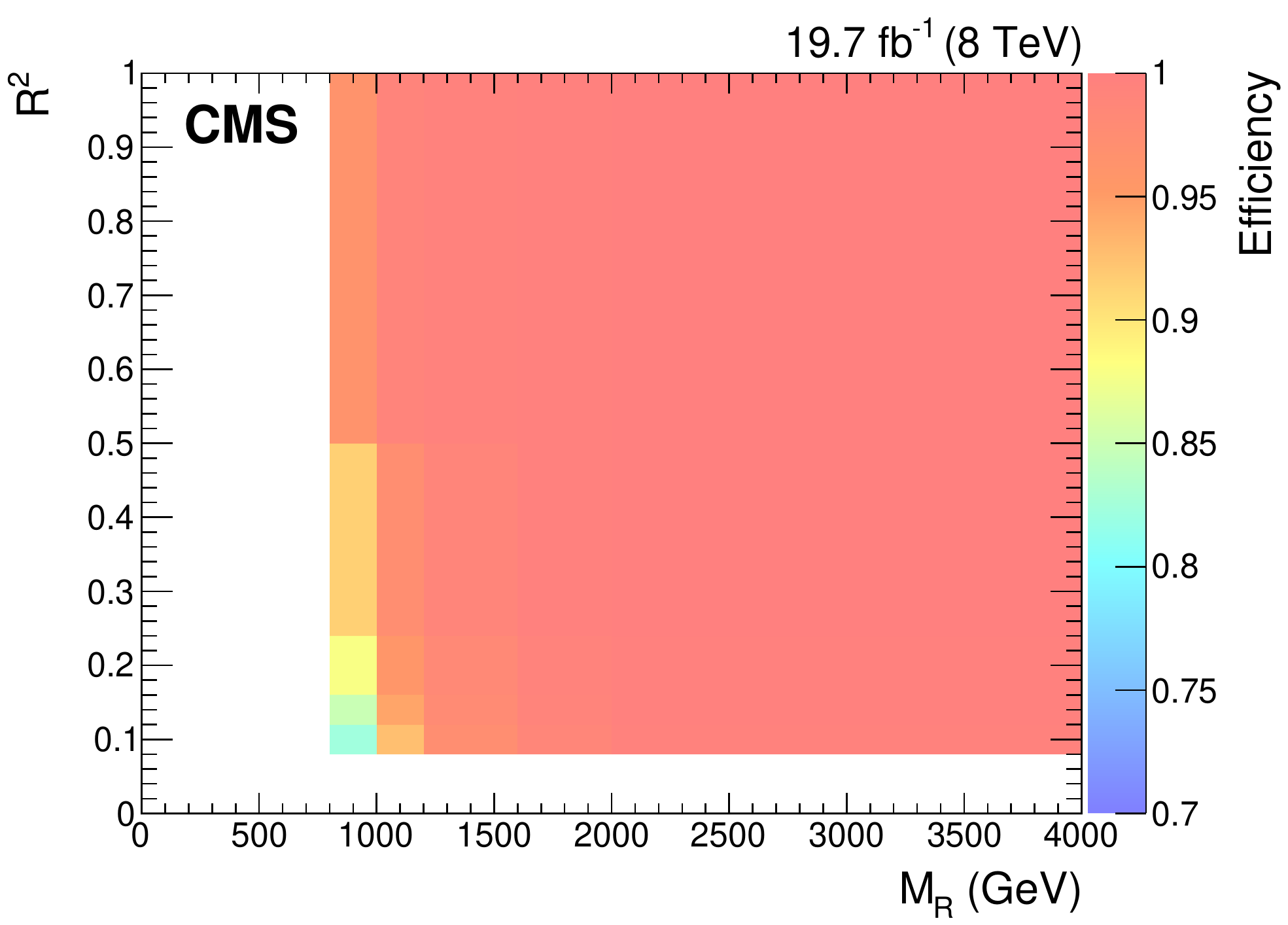}
\caption{(Left panel) The trigger efficiency, obtained from data, as a function of $\HT$ and leading jet $\pt$ after the baseline selection discussed in Section~\ref{sec:selection}. (Right panel) The trigger efficiency as a function of $\MR$ and $R^2$ after the same baseline selection, obtained by applying the trigger efficiency as a function of $\HT$ and leading jet $\pt$ to the simulated background.
\label{fig:trigger_efficiency}}
\end{figure*}

To measure the efficiency of these triggers, samples with unbiased jet \pt and $\HT$ distributions are obtained using an independent set of triggers that require at least one electron or muon.
Figure~\ref{fig:trigger_efficiency} shows, on the left-hand side, the efficiency of
the requirement that events satisfy at least one of the two trigger conditions as well as the baseline selection described in Section~\ref{sec:selection}, in the ($\HT$, leading jet \pt) plane.   The trigger is fully efficient for events with $\HT > 800\GeV$.  In order to account for the lower efficiency of the regions with $\HT  < 800\GeV$, the measured trigger efficiency over the ($\HT$, leading jet \pt) plane is applied as an event-by-event weight to the simulated samples. The right-hand side of Fig.~\ref{fig:trigger_efficiency} shows the trigger efficiency across the ($\MR,R^2$) plane for the total simulated background.

Simulated event samples are used to investigate the characteristics of the background and signal processes.  Multijet, $\cPqt\cPaqt$, $\PW({\to}\,\ell\PGn)+$jets, $\cPZ/\cPgg^*({\to}\,\ell\bar{\ell})+$jets, and $\cPZ({\to}\,\PGn\PAGn)+$jets events are generated using \MADGRAPH 5.1.3.30~\cite{Alwall:2011uj,MadGraph} with CTEQ6L1~\cite{Pumplin:2002vw} parton distribution functions (PDFs),
while $\PW\PW$, $\PW\cPZ$, and $\cPZ\cPZ$ events are generated using {\PYTHIA}6.424~\cite{Sjostrand:2006za} with CTEQ6L1 PDFs. In what follows, $\PW$ and $\cPZ$ bosons will be collectively referred to as $V$.
Single top quark events are generated using \POWHEG 1.0~\cite{powheg,powheg2} and CT10 PDFs~\cite{Lai:2010vv}.
The cross sections for these SM processes are given in Table~\ref{tab:cutflow}.
The inclusive background processes are scaled to the highest-order cross section calculation available, whereas leading-order cross sections are used for $\PW({\to}\,\ell\PGn)+$jets, $\cPZ/\cPgg^*({\to}\,\ell\bar{\ell})+$jets, and $\cPZ({\to}\,\PGn\PAGn)+$jets, which are produced with varying generator-level \HT requirements.
The simplified model signals are produced using \MADGRAPH 5.1.5.4 using CTEQ6L1 PDFs.  The signal cross sections are computed at next-to-leading order with next-to-leading-log corrections using \PROSPINO and \textsc{nll-fast}~\cite{Kramer:2012bx,NLONLL1,NLONLL2,NLONLL3,NLONLL4,NLONLL5}.
The parton-level events are showered and hadronized using {\PYTHIA}6.426 with tune Z2*~\cite{Chatrchyan:2013gfi}, which is derived from the Z1 tune~\cite{Field:2010bc}. The latter uses the CTEQ5L PDFs~\cite{Lai:1999wy}, whereas Z2* adopts CTEQ6L.
For the background events, the response of the CMS detector is simulated in detail using a program (\textsc{FullSim}) based on \GEANTfour~\cite{G4}.  A parametrized fast detector simulation program (\textsc{FastSim}) is used to simulate the detector response
for the signal events~\cite{fastsim}.

\section{Event reconstruction \label{sec:eventreco}}

We select events that have at least one interaction vertex associated with at least four charged-particle tracks. The vertex position is required to lie within 24\unit{cm} of the center of the CMS detector along the beam direction and within 2\unit{cm} from the center in the plane transverse to the beam.  Because of the high instantaneous luminosity of the LHC, hard scattering events are typically accompanied by overlapping events from multiple proton-proton interactions (pileup),
and therefore contain multiple vertices.
We identify the primary vertex, \ie, the vertex of the hard scatter, as the one with the highest value of the
$\sum \pt^2$ of the associated tracks.  Detector- and beam-related filters are used to discard events with anomalous noise that mimic events with high energy and a large imbalance in transverse momentum~\cite{Chatrchyan:2011tn, MET8TeV}.

CMS reconstructs events using the PF algorithm, in which
candidate particles (PF candidates) are formed by combining information from the inner tracker, the calorimeters, and the muon system.  Each PF candidate is assigned to one of five object categories: muons, electrons, photons, charged hadrons, and neutral hadrons.  Contamination from pileup
events is reduced by discarding charged PF candidates that are incompatible with having
originated from the primary vertex~\cite{CMS-PAS-JME-14-001}.   The average pileup energy associated with neutral hadrons is computed event by event and subtracted from the jet energy and from the energy used when computing lepton isolation, \ie, a measure of the activity around the lepton.  The energy subtracted is the average pileup energy per unit area (in $\Delta\eta \times \Delta\phi$) times the jet or isolation cone area~\cite{Fastjet1, Fastjet2}.

Jets are clustered with \textsc{FastJet 3.0.1}~\cite{Cacciari:2011ma} using the anti-\kt algorithm~\cite{antikt} with distance parameter $\Delta R=0.5$. These jets are referred to as AK5 jets.  Corrections are applied as a function of jet $\pt$ and $\eta$ to account for the residual effects of a nonuniform detector response.  The  jet energies are corrected so that, on average, they match those of simulated particle-level jets~\cite{Chatrchyan:2011ds}.  After correction, jets are required to have $\pt > 30\GeV$ and $\abs{\eta} < 2.4$.  We use the combined secondary vertex algorithm~\cite{btag7TeV,btag8TeV} to identify jets arising from $\cPqb$ quarks.  The medium tagging criterion, which yields a misidentification rate for light quark and gluon jets of ${\approx}1\%$ and a typical efficiency of ${\approx}70\%$, is used to select $\cPqb$ jets. The loose tagging criterion, with a misidentification rate of ${\approx}10\%$ and an efficiency of ${\approx}85\%$, is used to reject events containing $\cPqb$ jets.

To identify boosted $\PW$ bosons, we follow a similar procedure as outlined in Ref.~\cite{EXO-12-024}.
Jets are clustered with \textsc{FastJet} using the Cambridge-Aachen algorithm~\cite{Dokshitzer:1997in} and a distance parameter of 0.8, yielding CA8 jets.  Jet energy corrections for these jets are derived from the anti-\kt jets with distance parameter $\Delta R=0.7$. Simulations show that the corrections are valid for CA8 jets and have an additional uncertainty
$\leq 2$\%.

The jet mass is calculated from the constituents of the jet after jet pruning, which removes the softest constituents of the jet.  During jet pruning, the jet constituents are reclustered, and at each step the softer and larger-angle ``protojet'' of the two protojets to be merged is removed should it fail certain criteria~\cite{Ellis:2009su,Ellis:2009me}.
A CMS study has shown that jet pruning reduces pileup effects and provides good discrimination between boosted $\PW$ jets and quark/gluon ($\PQq$/$\Pg$) jets~\cite{Chatrchyan:2013vbb}.
We define mass-tagged jets ($mW$) as CA8 jets with $\pt > 200\GeV$ and jet mass within the range $70 < m_\text{jet} < 100\GeV$ around the $\PW$ boson mass.

In addition to the jet mass, we also consider the N-subjettiness~\cite{Thaler:2010tr} variables, which are obtained by first finding $N$ candidate axes for subjets in a given CA8 jet, and then computing the quantity
\begin{equation}
\ifthenelse{\boolean{cms@external}}
{
\begin{split}
\tau_N = \ \ \ & \\  \frac{1}{R_0} \sum_k  & p_{\mathrm{T}, k} \min (\Delta R_{1,k}, \Delta R_{2,k}, \cdots \Delta R_{N,k}) /
\sum_{k} p_{\mathrm{T}, k},
\end{split}
}
{
\tau_N = \frac{1}{R_0} \sum_k p_{\mathrm{T}, k} \min (\Delta R_{1,k}, \Delta R_{2,k}, \cdots \Delta R_{N,k}) /
\sum_{k} p_{\mathrm{T}, k},
}
\end{equation}
where $R_0$ is the original jet distance parameter and $k$ runs over all constituent particles.
The subjet axes are obtained with \textsc{FastJet} via exclusive $k_\textrm{T}$ clustering,
followed by a one-pass optimization to minimize the N-subjettiness value.
The quantity $\tau_N$ is small if the original jet is consistent with having $N$ or fewer subjets.
Therefore, to discriminate boosted $\PW$ bosons, which have two subjets, from $\PQq$/$\Pg$ jets characterized by a single subjet, we require that a $\PW$ boson mass-tagged jet satisfy $\tau_2 / \tau_1 < 0.5$ for it to be classified as a $\PW$ boson tagged jet (labeled $\PW$ in the following).
The $\PW$ boson tagging efficiency is dependent on the CA8 jet \pt, and is 50--55\% according to simulation.  The corresponding misidentification rate is 3--5\%.
We also define $\PW$ boson antitagged jets ($aW$) as $\PW$ boson mass-tagged jets that satisfy the complement of the $\tau_2 / \tau_1$ criterion, and use these jets to define control regions for data-driven background modeling.

To calculate \ptvecmiss, which is used in the calculation of the razor variable $R^2$ defined in Eqs.~(\ref{eq:MRT}) and (\ref{eq:R2}), the vector sum over the transverse momenta is taken of all the PF candidates in an event.

Loosely identified and isolated electrons~\cite{Khachatryan:2015hwa} (and muons~\cite{Chatrchyan:2012xi}) with $\pt > 5\GeV$ and $\abs{\eta} < 2.5$ ($2.4$) are used both to suppress backgrounds in the signal region and in the definition of the control regions.
Tightly identified isolated leptons, electrons (muons) with $\pt > 10\GeV$ and $\abs{\eta} < 2.5$ ($2.4$), define a control region enriched in $\cPZ{\to}\,\ell \bar{\ell}$ events, from which we estimate the systematic uncertainty in the predicted number of $\cPZ {\to}\, \PGn \PAGn$ events in the signal region.
Electron candidates that lie in the less well-instrumented transition region between the barrel and end cap calorimeters, $1.44 < \abs{\eta} < 1.57$, are rejected.
We suppress the background from events that are likely to contain $\PGt$ and other leptons that fail the loose selection by discarding events with isolated tracks with $\pt > 10\GeV$ and a track-primary vertex distance along the beam direction $|d_z| < 0.05\unit{cm}$.

Known differences between the properties of data and MC simulated data are corrected by weighting simulated events with data/simulation scale factors for
the jet energy scale, $\cPqb$ tag, $\PW$ mass-tag, $\PW$ tag, and $\PW$ antitag efficiency. The $\PW$ tagging-related scale factors are described in Section~\ref{sec:Wtag_SF}.
In addition, event-by-event weights are used to correct the simulated data so that
their pileup, trigger, top quark $\pt$, and ISR characteristics match those of the data.

\section{Analysis strategy and event selection
\label{sec:selection}}
We search for deviations from the SM in the (high-$\MR$, high-$R^2$) region
using events with at least one boosted $\PW$ boson, at least one $\cPqb$-tagged jet, and no isolated leptons or tracks.  SM backgrounds in the signal region $S$ are estimated using observations in control regions and scale factors, calculated from MC simulation, that relate the number of events in one region to that in another.
Three control regions, $Q$, $T$, and $W$, select high-purity samples of multijet, $\cPqt\cPaqt$, and $\PW({\to}\,\ell\PGn)+$jets events, respectively.  Details of the background estimation method are given in Section~\ref{sec:likelihood}.

Events must satisfy the following baseline selection:
\begin{enumerate}
 \item have at least one good primary vertex (see Section~\ref{sec:eventreco});
 \item pass all detector- and beam-related filters (see Section~\ref{sec:eventreco});
 \item have at least three selected AK5 jets of which at least one has  $\pt > 200\GeV$, thereby
 defining the boosted phase space; and
 \item satisfy $\MR > 800\GeV$ and $R^2 > 0.08$, where the megajets are constructed from the selected AK5 jets.
\end{enumerate}

The details of the event selection in addition to the baseline selection are given in Table~\ref{tab:selection}.
The signal and control regions are defined using different requirements on the
multiplicities of leptons, $\cPqb$-tagged jets, and $\PW$-tagged jets, and on kinematic variables that discriminate between different processes.
The multijet-enriched control sample $Q$ is used for estimating the multijet background in the $S$ and $T$ regions.  To characterize $Q$, we use the fact that \ETm in multijet events is largely due to jet mismeasurements rather than the escape of particles that interact weakly with the detector;  consequently, \ptvecmiss will often be aligned with one of the jets.  Therefore, a good discriminant between multijet events and events with genuine \ETm  is
\begin{equation}
 \Delta\phi_\text{min} = \min_i{\Delta\phi(\ptvecmiss, {\vec p}_{\mathrm{T}\, i}   ) },
\end{equation}
that is, the minimum of the angles between \ptvecmiss and the transverse momentum of each jet,
where $i$ runs over the three leading AK5 jets. Since detector inaccuracies mostly cause undermeasurements of the jet energy and momentum, the variable $\Delta\phi_\text{min}$ provides a reliable discrimination of fake $\ETm$ in multijet events.

\begin{table*}[htbp]
\centering
\topcaption{Summary of the selections used, in addition to the baseline selection, to define the signal region ($S$), the three control regions ($Q$, $T$, $W$), and the two regions ($S'$, $Q'$) used for the cross-checks described later in the text.
\label{tab:selection}}
\begin{scotch}{lcccccc}
Selection & $S$ & $S'$ & $Q$ & $Q'$& $T$  & $W$ \\
\hline
Number of $\cPqb$-tagged jets & ${\geq} 1$   & ${\geq} 1$   & 0   & 0     & ${\geq} 1$      & 0 \\
Number of mass-tagged $\PW$s  & ${\geq} 1$   & ${\geq} 1$   & ${\geq} 1$ & ${\geq} 1$ & ${\geq} 1$      & ${\geq} 1$ \\
Number of tagged $\PW$s       & ${\geq} 1$   & ${\geq} 1$   & \NA   & \NA    & ${\geq} 1$      & \NA \\
Number of antitagged $\PW$s  & \NA          & \NA          & ${\geq} 1$ & ${\geq} 1$ & \NA             & \NA \\
Number of loose leptons       & 0          & 0          & 0    & 0    & 1             & 1 \\
Number of isolated tracks     & 0          & 0          & 0   & 0     & \NA             & \NA \\
$m_\mathrm{T}$ (\GeVns)            & \NA          & \NA          & \NA   & \NA     & ${<} 100$   & 30--100\\
$\Delta\phi_\text{min}$        & ${>} 0.5$    & ${<} 0.5$    & ${<} 0.3$  & ${>} 0.5$  & ${>} 0.5$       & ${>} 0.5$\\
\end{scotch}
\end{table*}

The $T$ and $W$ control regions are used to characterize the $\cPqt\cPaqt$ and $\PW+$jets backgrounds, respectively, in the $S$ region.
The contamination in the $S$ region from fully hadronic decays of $\cPqt\cPaqt$ pairs is negligible because they
do not produce sufficient genuine $\ETm$ to satisfy our event selection.
The $\cPqt\cPaqt$ contamination
consists thus of the semileptonic decays of $\cPqt\cPaqt$ pairs in which
one $\PW$ boson is boosted and the other $\PW$ boson decays
to a charged lepton that is not identified.
Therefore, the $T$ region is required to have a lepton from the decay of a $\PW$ boson, at least
one $\cPqb$-tagged jet, and a $\PW$-tagged jet.
Similarly, the $\PW+$jets contribution in the $S$ region  comes from leptonic $\PW$ boson decays in which the charged lepton is not identified and a jet is misidentified as a $\PW$ jet.  Therefore, we require the $W$ region to have events with a lepton from the $\PW$ boson and a mass-tagged boosted $\PW$ jet, which is a quark or gluon initiated jet misidentified as a boosted $\PW$ boson.  The N-subjettiness criterion is not imposed in order to maintain high event yields in these control regions and
therefore higher statistical precision.

In the $T$ and $W$ regions, we suppress potential signals using the transverse mass,
\begin{equation}
 m_\mathrm{T} = \sqrt{2\pt^\ell\ETm ( 1 - \cos\Delta\phi )},
\end{equation}
where $\Delta\phi$ is the difference in azimuthal angle between the lepton \ptvec and \ptvecmiss, and $\pt^\ell$ is the magnitude of the lepton \ptvec.
The $m_\mathrm{T}$ distribution exhibits a kinematic edge at the mass of the $\PW$ boson for $\cPqt\cPaqt$ and $\PW({\to}\ell\PGn)+$jets processes. However, such an edge is not present for signal events because of the extra contribution to \ETm from neutralinos, which escape direct detection.  Therefore, potential signals are suppressed in the $T$ and $W$ regions by requiring $m_\mathrm{T} < 100$\GeV.  For the $W$ region, we additionally require $m_\mathrm{T} > 30$\GeV in order to reduce residual contamination from multijet events, which are expected to have small \ETm and therefore small $m_\mathrm{T}$.   Table~\ref{tab:selection} lists two additional control regions, $S'$ and $Q'$, which are used in the
cross-checks described later in this section.

\begin{figure*}[htb]
 \includegraphics[width=0.49\textwidth]{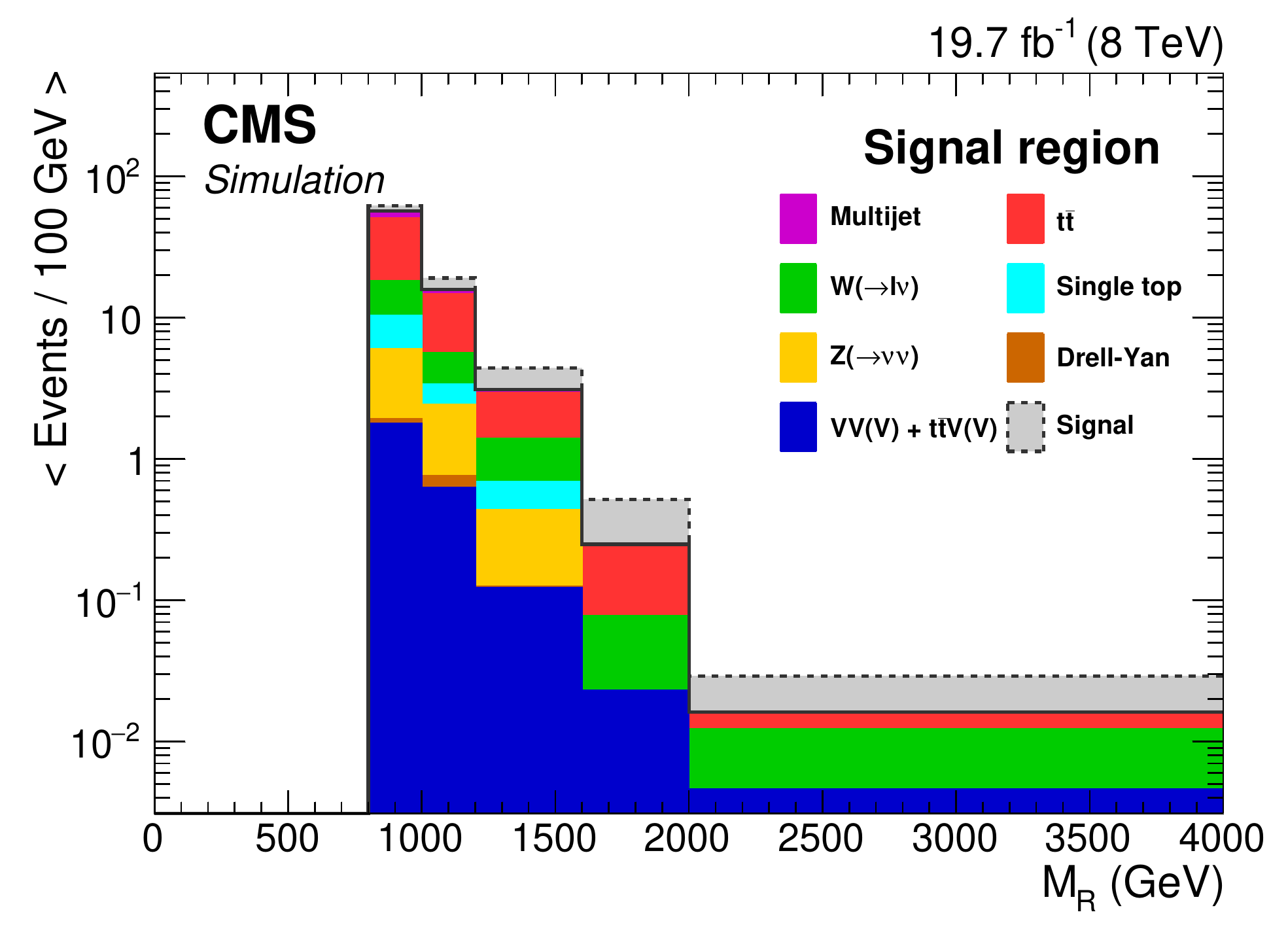}
 \includegraphics[width=0.49\textwidth]{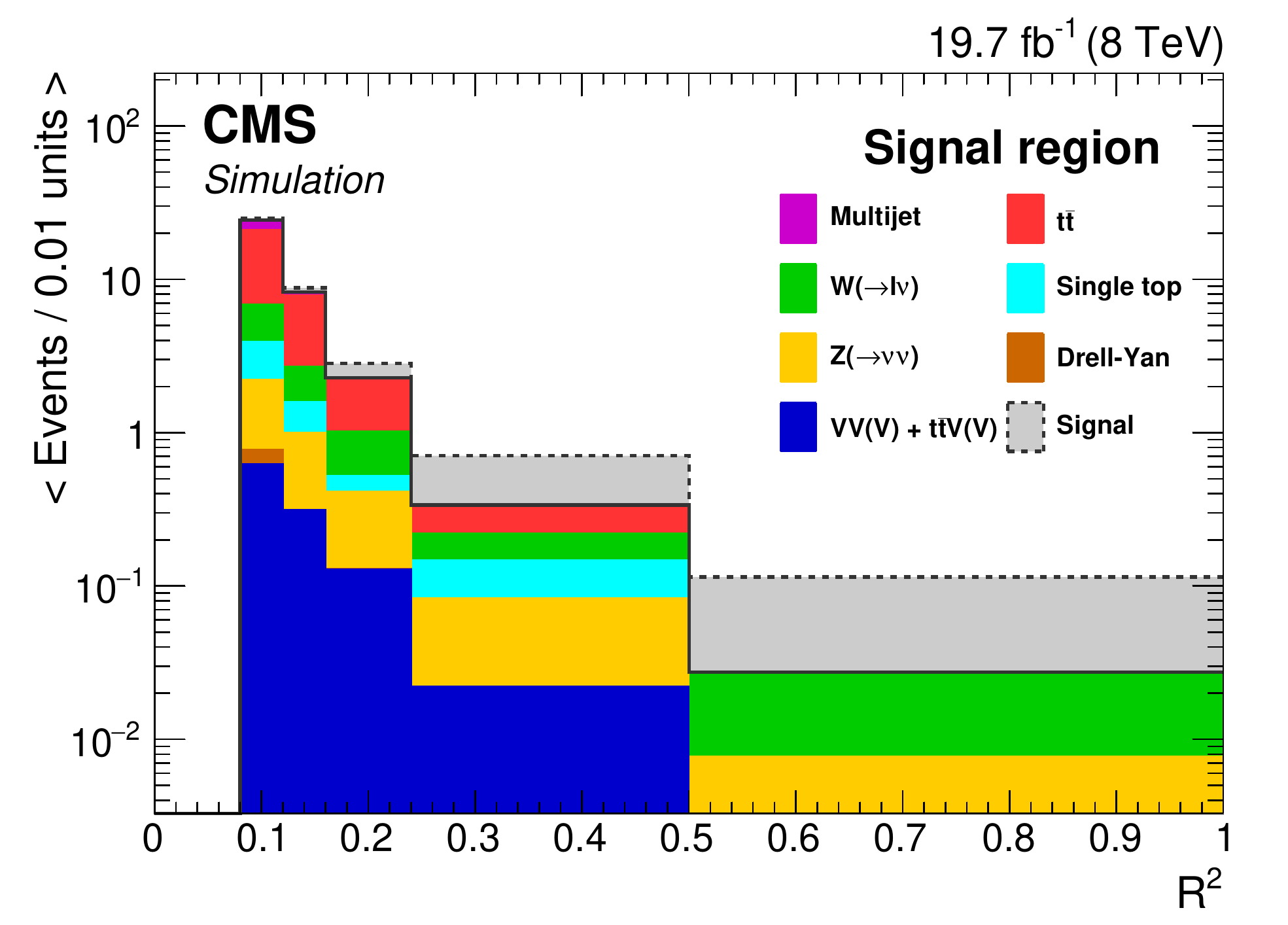} \\
\caption{Simulated $\MR$ (left panel) and $R^2$ (right panel) distributions in the signal region, $S$.
Stacked on top of the background distributions is
the predicted signal contribution from an example T1ttcc model, with parameters  $m_{\PSg} =1\TeV$, $m_{\PSQt} =325\GeV$, and $m_{\PSGczDo} =300\GeV$. The bin entries are normalized proportionally to the bin width.
\label{fig:DataMC}}
\end{figure*}

\begin{figure*}[htb]
\centering
 \includegraphics[width=0.49\textwidth]{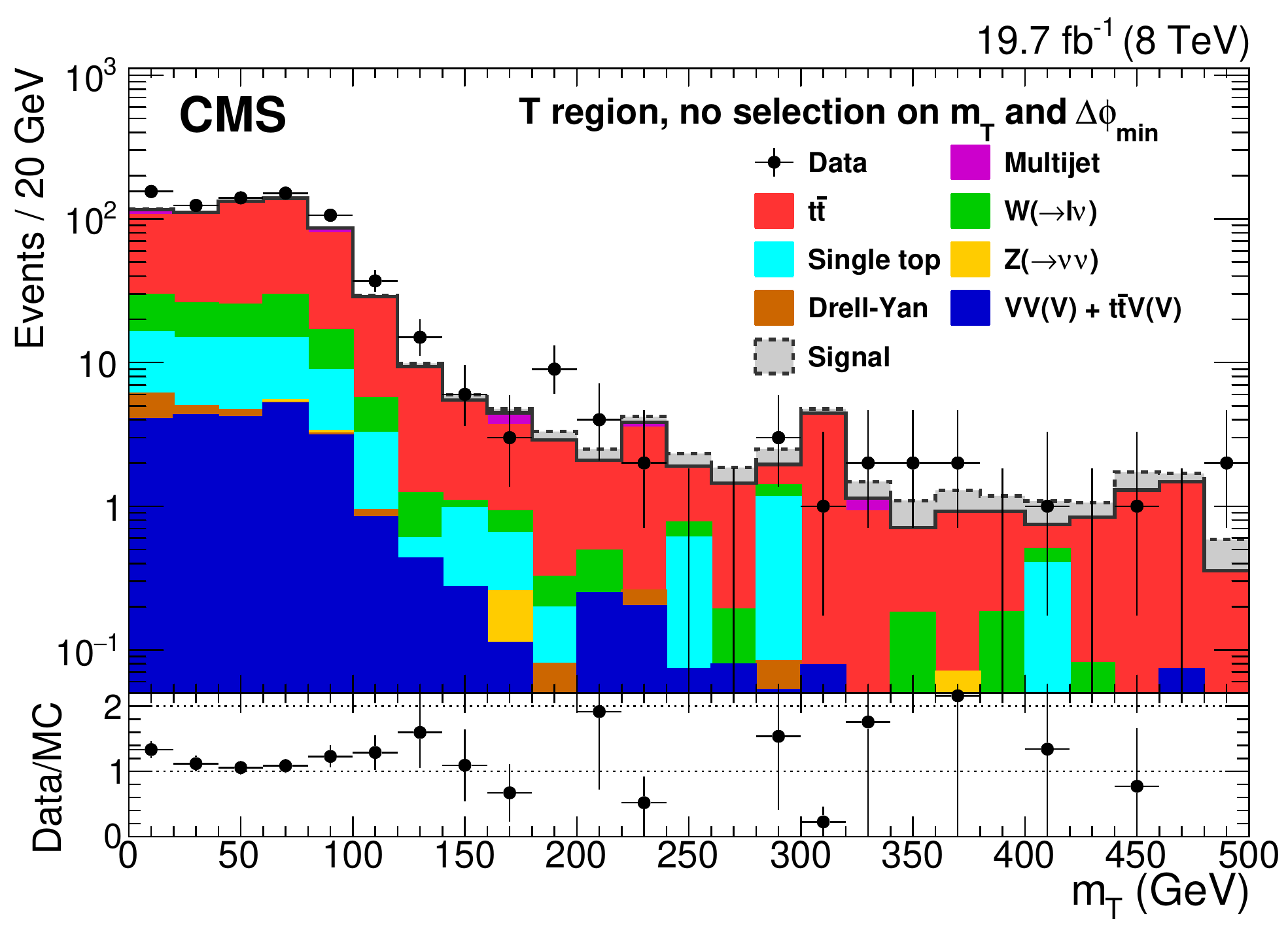}
 \includegraphics[width=0.49\textwidth]{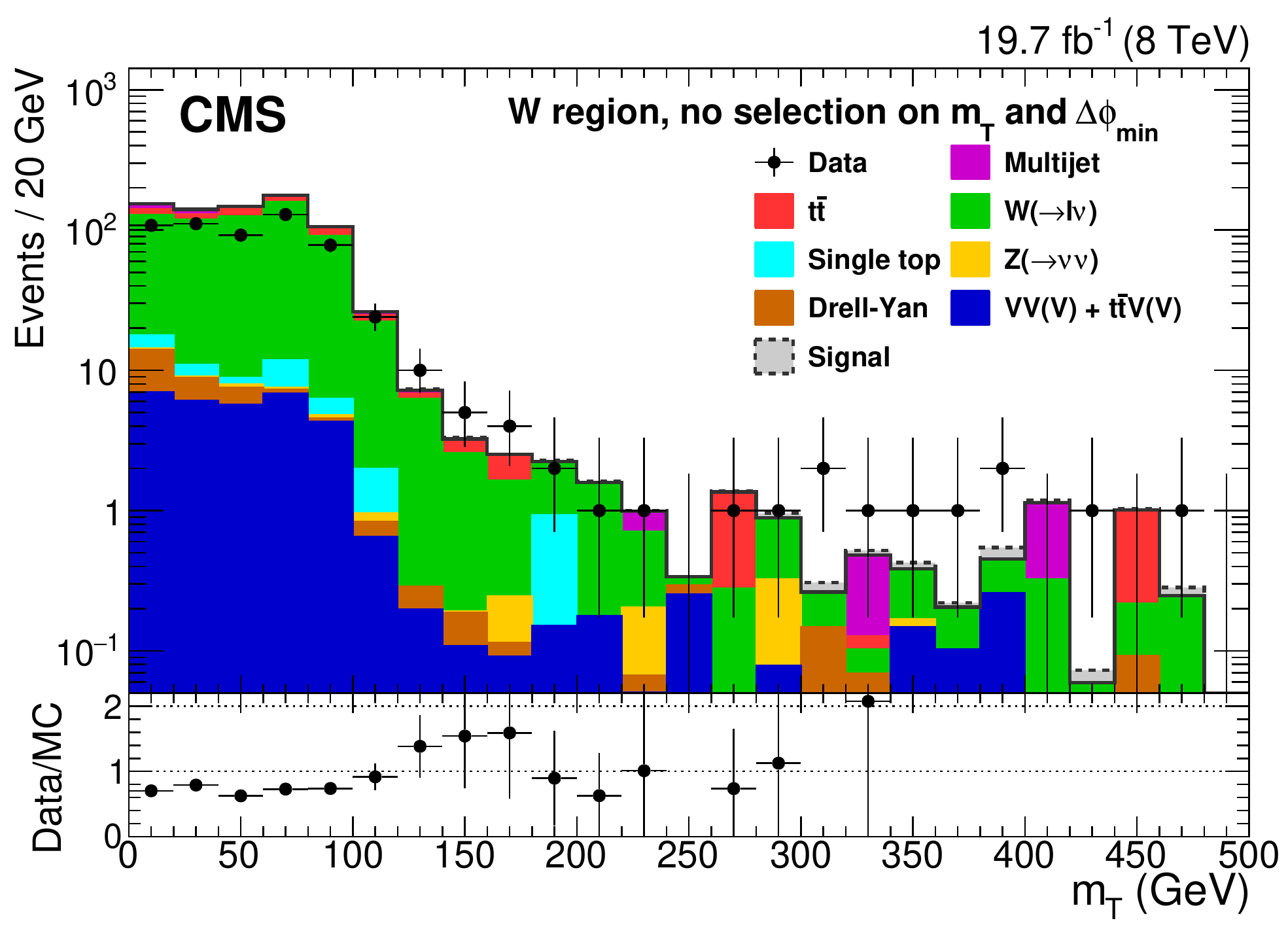}
\caption{Distributions of $m_\mathrm{T}$ for data and simulated backgrounds, in the $T$ (left panel) and $W$ (right panel) control regions, without applying any selection on $m_\mathrm{T}$ and $\Delta\phi_\text{min}$. The contribution from an example signal corresponding to the T1ttcc model with $m_{\PSg} =1\TeV$, $m_{\PSQt} =325\GeV$, and $m_{\PSGczDo} =300\GeV$, is stacked on top of the background processes. Only statistical uncertainties are shown.
\label{fig:mT}}
\end{figure*}

\begin{figure}[htb]
\centering
 \includegraphics[width=0.45\textwidth]{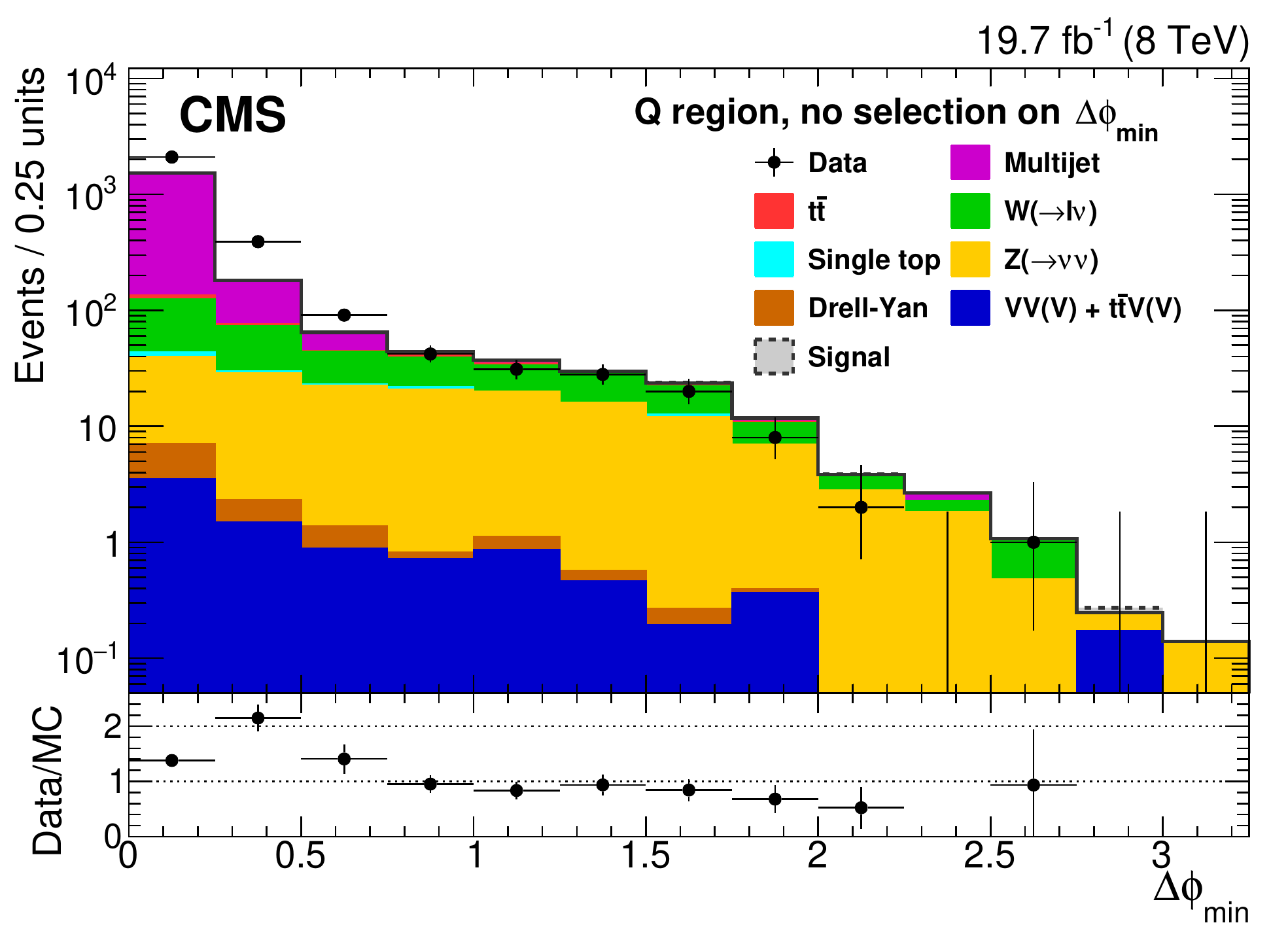}
\caption{Distributions of $\Delta\phi_\text{min}$ for data and simulated backgrounds in the $Q$ region without applying a selection on $\Delta\phi_\text{min}$. Only statistical uncertainties are shown. Signal contamination in this control region is negligible.
\label{fig:deltaphi}}
\end{figure}

Figure~\ref{fig:DataMC} shows the simulated distributions in the signal region for the $\MR$ and $R^2$ variables, where the smoothly falling nature of the backgrounds, as well as their relative contributions, can be observed.
The $m_\mathrm{T}$ distribution in the $T$ and $W$ regions prior to the $m_\mathrm{T}$ and $\Delta\phi_\text{min}$
selection is shown in Fig.~\ref{fig:mT}, while Fig.~\ref{fig:deltaphi} shows the $\Delta\phi_\text{min}$ distribution in the $Q$ region, for both data and simulated backgrounds.
Overall, there is reasonable agreement between the
observed and simulated yields.  The discrepancies are accommodated by the systematic uncertainties we assign to the simulated yields.

In Table~\ref{tab:cutflow}, we show the expected number of events obtained from simulation for the different background processes and for the example T1ttcc model with $m_{\PSg} = 1\TeV$, $m_{\PSQt} = 325\GeV$, and $m_{\PSGczDo} = 300\GeV$.
The observed event counts after different levels of selection, beyond the trigger requirement, are also reported.
The background composition in percent after the baseline, $S$, $Q$, $T$, and $W$ region selections is reported in Table~\ref{tab:BG_comp_percent}.
The signal region is $\cPqt\cPaqt$ dominated, with additional contributions from $\PW({\to}\,\ell\PGn)+$jets and multijet processes. Each control region, $Q$, $T$, and $W$, has
high purity for the background process it targets, 90\% multijet, 83\% $\cPqt\cPaqt$ and single top quark processes, and 85\% $\PW({\to}\,\ell\PGn)+$jets, respectively.  The discrepancies between the observations and the simulation are due to uncertainties in the MC modeling, especially for the multijet processes.

\begin{table*}[p]
\centering
\topcaption{Event yields in simulated event samples and in data as event selection requirements are applied.
The simulated event counts are normalized to an integrated luminosity of $19.7\fbinv$. ``Other'' refers to the sum of the small
background components $\cPZ/\cPgg^*{\to}\,\ell\bar{\ell}+$jets, triboson, and $\cPqt\cPaqt V$.
The signal is the T1ttcc model with $m_{\PSg}=1000\GeV$, $m_{\PSQt_1}=325\GeV$, $m_{\PSGczDo}=300\GeV$.
The row corresponding to ``$n_\mathrm{PV} > 0$'' gives the event counts after applying the noise filters, pileup reweighting, top \pt reweighting for $\cPqt\cPaqt$, ISR reweighting for the signal, and the requirement of at least one primary vertex. The column listing the total number of background events also includes some processes that only contribute at the early stages of the event selection.
The cross sections used for each sample are listed in the second line of the header. Several of the simulated background samples were produced with generator-level selections applied, which are not fully covered by the first selection levels listed in this table.
}
\renewcommand{\arraystretch}{1.1}
\cmsTable{
\begin{scotch}{ l | c  c  c  c  c  c  c | c | c | c }
\multirow{2}{*}{Selection} & Multijet & $\cPqt\cPaqt$ & $\PW({\to}\ell\PGn)$ & Diboson & Single top & $\cPZ({\to}\PGn\PAGn)$ & Other & Total & Signal & \multirow{2}{*}{Data} \\
 & $10.4{\times}10^7$ pb & 245.8 pb & 111.5 pb & 95.4 pb & 114.9 pb & 588.3 pb & 25.2 pb & background & 0.02435 pb & \\ \hline \hline
No selection & $\phantom{0}2.1{\times}10^{11}$ & $\phantom{0}4.9{\times}10^6$ & $\phantom{0}2.2{\times}10^6$ & $\phantom{0}1.9{\times}10^6$ & $\phantom{0}2.3{\times}10^6$ & $\phantom{0}1.2{\times}10^7$ & $\phantom{0}4.9{\times}10^5$ & $\phantom{0}2.1{\times}10^{11}$ & 499 &  \\
$n_\mathrm{PV} > 0$ & $1.05{\times}10^{11}$ & $4.42{\times}10^6$ & $2.02{\times}10^6$ & $1.08{\times}10^6$ & $1.72{\times}10^6$ & $2.87{\times}10^6$ & $\phantom{0}4.0{\times}10^5$ & $1.05{\times}10^{11}$ & 479 & \\
$n_\mathrm{j} \geq 3$ & $2.04{\times}10^{10}$ & $4.08{\times}10^6$ & $1.51{\times}10^6$ & $5.19{\times}10^5$ & $1.10{\times}10^6$ & $6.24{\times}10^5$ & $3.37{\times}10^5$ & $2.05{\times}10^{10}$ & 472 &  \\
$\pt(\rm j_1) > 200\GeV$ & $1.82{\times}10^{8\phantom{0}}$ & $2.88{\times}10^5$ & $4.36{\times}10^5$ & $1.86{\times}10^4$ & $6.08{\times}10^4$ & $5.89{\times}10^{4}$ & $7.23{\times}10^4$ & $1.82{\times}10^{8\phantom{0}}$ & 403 & \\
$\MR \,{>}\, 800, R^2 \,{>}\, 0.08$ & $3.47{\times}10^{4\phantom{0}}$ & $5.83{\times}10^3$ & $1.17{\times}10^4$ & 309 & 900 & $3.25{\times}10^3$ & 645 & $57\,557$ & 224 & \\
Trigger & $3.15{\times}10^{4\phantom{0}}$ & $5.12{\times}10^3$ & $9.38{\times}10^3$ & 249 & 786 & $2.32{\times}10^3$ & 569 & $50\,164$ & 216 & $67\,037$ \\
\hline \hline
No leptons & $3.09{\times}10^{4\phantom{0}}$ & $1.87{\times}10^3$ & $3.75{\times}10^3$ & 96.3 & 311 & $2.30{\times}10^3$ & 216 & $39\,666$ & 142 & $56\,220$ \\
\hline
$n_\cPqb \geq 1$ & $9.37{\times}10^{3\phantom{0}}$ & $1.51{\times}10^3$ & 590 & 25.2 & 226 & 302 & 79.8 & $12\,187$ & 119 & $18\,164$ \\
$n_\PW \geq 1$ & 841 & 332 & 56.4 & 8.52 & 56.7 & 22.1 & 16.9 & $1\,350$ & 28 & $1\,817$  \\
$S$ & 14.8 & 90.4 & 23.1 & 3.7 & 11.7 & 12.7 & 4.17 & 160 & 23.4 & 187 \\
\hline
$n_\cPqb = 0$ & $1.25{\times}10^{4\phantom{0}}$ & 98.3 & $1.70{\times}10^3$ & 35.6 & 25.9 & $1.25{\times}10^3$ & 54.3 & $15\,691$ & 5.65 & $20\,667$ \\
$n_\mathrm{aW} \geq 1$ & 1519 & 18.7 & 204 & 8.36 & 7.40 & 158 & 6.98 & $1\,923$ & 0.667 & $2\,712$ \\
$Q$ & 1447 & 10.6 & 93.1 & 3.88 & 3.94 & 38.9 & 4.48 & $1\,603$ & 0.07 & $2\,240$ \\
\hline
\hline
1 lepton & 585.9 & $2.74{\times}10^3$ & $5.52{\times}10^3$ & 132 & 421 & 22.1 & 272 & $9\,699$ & 65.0 & $10\,008$ \\
\hline
$n_\cPqb \geq 1$ & 236.7 & $2.17{\times}10^3$ & 625 & 29.9 & 301 & 4.14 & 102 & $3\,470$ & 54 & $3\,930$ \\
$n_\PW \geq 1$ & 24.3 & 496 & 61.6 & 10.0 & 50.9 & 0.56 & 21.9 & 666 & 12.3 & 770 \\
$T$ & 0 & 112 & 20.2 & 2.0 & 13.3 & 0 & 4.1 & 151 & 1.2 & 153 \\
\hline
$n_\cPqb = 0$ & 150.5 & 153 & $2.86{\times}10^3$ & 52.8 & 41.3 & 11.5 & 68.8 & $3\,329$ & 2.54 & $3\,165$ \\
$n_{mW} \geq 1$ & 30.8 & 79.1 & 605 & 33.1 & 13.8 & 2.4 & 20.3 & 786 & 1.19 & 581 \\
$W$ & 0 & 15.5 & 127 & 3.6 & 1.6 & 0.64 & 1.4 & 150 & 0.06 & 116 \\
\end{scotch}
}
\label{tab:cutflow}
\end{table*}

\begin{table*}[tpb]
\centering
\topcaption{Background composition according to simulation after the baseline, $S$, $Q$, $T$, $W$, $Q'$ and $S'$ region selections. ``Other'' refers to the sum of the small
background components $\cPZ/\cPgg^*{\to}\,\ell\bar{\ell}$, triboson, and $\cPqt\cPaqt V$.
\label{tab:BG_comp_percent}}
\newcolumntype{.}{D{.}{.}{-1}}
\begin{scotch}{ l  .  .  .  .  .  .  . }
\multirow{2}{*}{Selection} & \multicolumn{1}{c}{Multijet} & \multicolumn{1}{c}{ $\cPqt\cPaqt$}  & \multicolumn{1}{c}{$\PW({\to}\,\ell\PGn)$} & \multicolumn{1}{c}{Diboson} & \multicolumn{1}{c}{Single top} & \multicolumn{1}{c}{$\cPZ({\to}\,\PGn\PAGn)$} & \multicolumn{1}{c}{Other} \\
& \multicolumn{1}{c}{(\%)} & \multicolumn{1}{c}{(\%)} & \multicolumn{1}{c}{(\%)} & \multicolumn{1}{c}{(\%)} & \multicolumn{1}{c}{(\%)} & \multicolumn{1}{c}{(\%)} & \multicolumn{1}{c}{(\%)} \\
\hline
Baseline & 62.8 & 10.2 & 18.7 & 0.5 & 1.6 & 4.6   & 1.6 \\
$S$      & 9.2  & 56.3 & 14.4 & 2.3 & 7.3 & 7.9   & 2.6 \\
$Q$      & 90.2 & 0.7  & 5.8  & 0.2 & 0.2 & 2.4   & 0.3 \\
$T$      & 0.0  & 73.9 & 13.3 & 1.3 & 8.8 & 0.0   & 2.7 \\
$W$      & 0.0  & 10.3 & 84.8 & 2.4 & 1.1 & 0.4   & 1.0 \\
$Q'$     & 12.3 & 2.8  & 36.8 & 1.7 & 1.0 & 45.0  & 0.4 \\
$S'$     & 69.5 & 20.3 & 2.8  & 0.4 & 3.8 & 0.8   & 2.4 \\
\end{scotch}
\end{table*}

We do not explicitly estimate the background in the signal region. Rather,
from the observations in the control regions,
we create a prior distribution (described in  Section~\ref{sec:likelihood})
for the four background components of the signal region that incorporates all
statistical and systematic uncertainties.  However,
in order to verify that the control regions in data provide adequate models for
backgrounds in the signal region and that the translations between different regions behave as expected, we perform two cross-checks, taking into account statistical uncertainties only.

In the first cross-check, we predict the background in a signal-like control region, and compare
these predictions with the observations in that region.
This control region, denoted by $S^\prime$, is defined by inverting the $\Delta\phi_\text{min}$ requirement while preserving the rest of the signal selection.
The estimated number of events in the $S^\prime$ region for the multijet, $\PW(\to\ell\PGn)+$jets, and top quark processes is computed as follows:
\begin{equation}
 \widehat{N}_\text{multijet}^{S^\prime} = \left( N_\text{obs}^{Q} - N_\text{other, MC}^{Q} \right)  /
                                    \left( \frac{N_\text{multijet}^{Q}}{N_\text{multijet}^{S^\prime}} \right)_\mathrm{MC},
\label{eq:E1}
\end{equation}
\begin{equation}
 \widehat{N}_{\PW(\to\ell\PGn)}^{S^\prime} = \left( N_\text{obs}^{W} - N_\text{other, MC}^{W} \right) /
                                       \left( \frac{N_{\PW(\to\ell\PGn)}^{W}}{N_{\PW(\to\ell\PGn)}^{S^\prime}} \right)_{\mathrm{MC}},
\label{eq:E2}
\end{equation}
\begin{equation}
\ifthenelse{\boolean{cms@external}}
{
\begin{split}
  \widehat{N}_\mathrm{TTJ+T}^{S^\prime} & =  \\
  \left( N_\text{obs}^{T} \right. & \left. - \widehat{N}_\text{multijet}^{T} - N_\text{other, MC}^{T} \right) /
  \left( \frac{N_\mathrm{TTJ+T}^{T}} {N_\mathrm{TTJ+T}^{S^\prime}}\right)_{\mathrm{MC}},
  \end{split}
}
{
  \widehat{N}_\mathrm{TTJ+T}^{S^\prime} = \left( N_\text{obs}^{T} - \widehat{N}_\text{multijet}^{T} - N_\text{other, MC}^{T} \right) /
                                       \left( \frac{N_\mathrm{TTJ+T}^{T}} {N_\mathrm{TTJ+T}^{S^\prime}}\right)_{\mathrm{MC}},
}
\label{eq:E3}
\end{equation}
while the estimated number of multijet events in the control region $T$ is given by
\begin{equation}
 \widehat{N}_\text{multijet}^{T} = \left( N_\text{obs}^{Q} - N_\text{other, MC}^{Q} \right) /
                             \left( \frac{N_\text{multijet}^{Q}}{N_\text{multijet}^{T}} \right)_{\mathrm{MC}}.
\label{eq:E4}
\end{equation}
In Eqs.~(\ref{eq:E1})--(\ref{eq:E4}), the superscripts denote one of the control regions, while the subscripts ``other", $\PW(\to\ell\PGn)$,
$\rm{TTJ+T}$, and multijet, denote the sum of the small backgrounds, $\PW(\to\ell\PGn)$+jets, $\cPqt\cPaqt$ plus
single top quark, and multijet, respectively, while ``obs" labels observed counts.
These equations are used only in this cross-check. However, they incorporate the same relations between signal and control regions as will be used in the likelihood procedure described in Section~\ref{sec:likelihood}.
As can be seen from Table~\ref{tab:BG_comp_percent}, the nominal choice of the parameters associated with systematic uncertainties leads to $N_\text{multijet, MC}^{T} = 0$.
The total estimated background in $S^\prime$ is
\begin{equation}
  \widehat{N}^{S^\prime} = \sum_i \widehat{N}^{S^\prime}_i ,
\end{equation}
where $i$ runs over all background processes.  For smaller backgrounds, $\widehat{N}^{S^\prime}_i$ is determined by simulation.
Backgrounds are estimated bin by bin in the $(\MR,R^2)$ space, where the bin boundaries are  numerically defined in Table~\ref{tab:results_prediction}.
However, the estimated scale factors are global as the statistical precision is not sufficient to yield reliable bin-by-bin estimates. The expected global scale factors, which we denote by $\kappa$,  are defined in Section~\ref{sec:likelihood}, which also describes how they are calculated.

\begin{figure*}[tpb]
\includegraphics[width=0.45\textwidth]{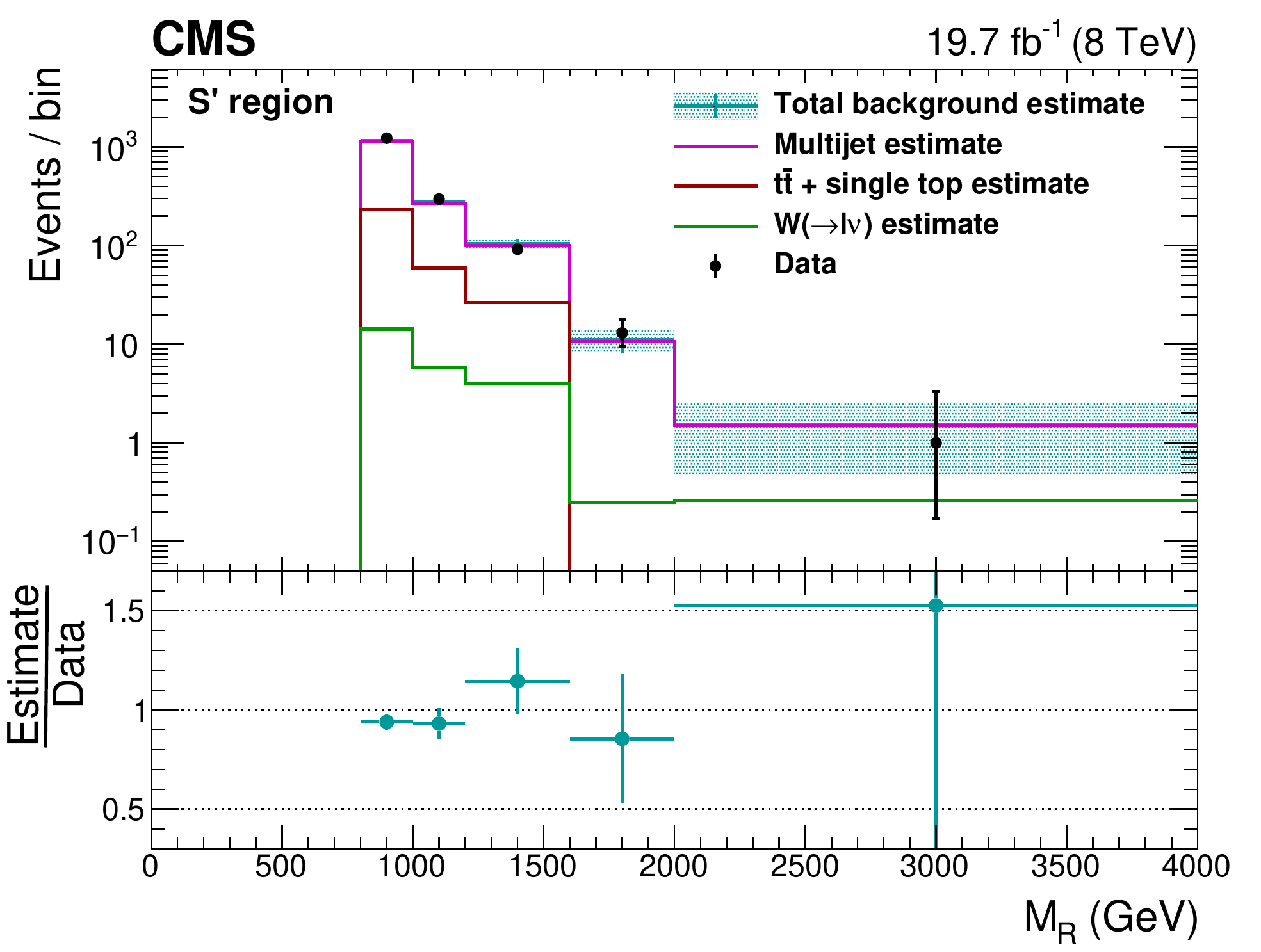}
\includegraphics[width=0.45\textwidth]{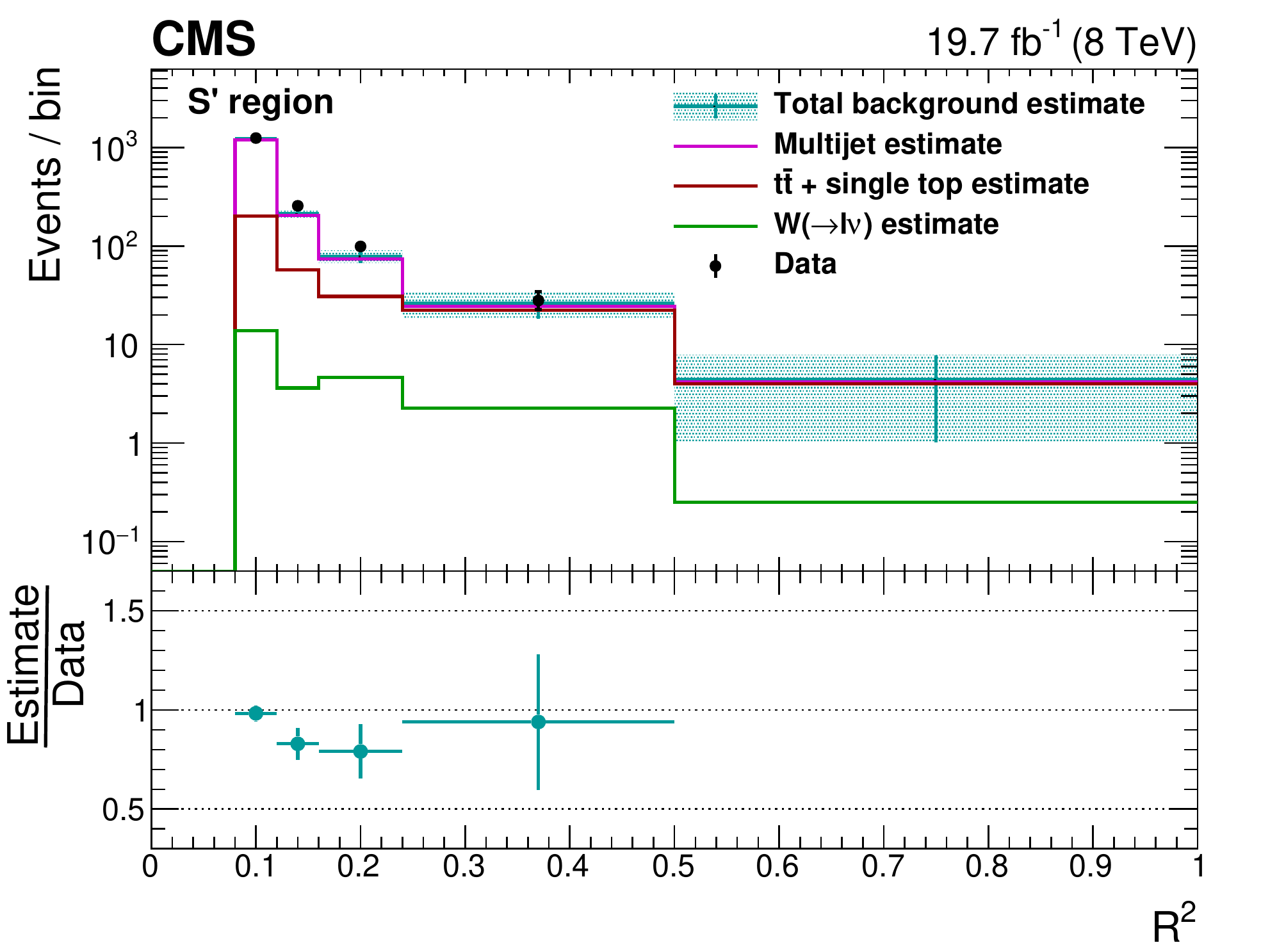}
\caption{One-dimensional projection of $\MR$ (left panel) and $R^2$ (right panel) for the cross-check predicting the $\Delta\phi_\text{min}$ sideband region $S'$.
The estimates for the three different background processes are stacked on top of each other.
The uncertainties shown are statistical only. The horizontal error bars indicate the bin width. \label{fig:Shape_syst_1D_project_sideband}}
\end{figure*}

Figure~\ref{fig:Shape_syst_1D_project_sideband} shows the projection on the $\MR$ and $R^2$ axes of the predicted and observed distributions in the $S'$ region.  The prediction agrees with  observation within ${\approx}20\%$.  This cross-check of the background modeling shows that it is feasible to estimate a multicomponent  background in a signal-like region using the control regions we
have defined.

\begin{figure*}[tpb]
\includegraphics[width=0.45\textwidth]{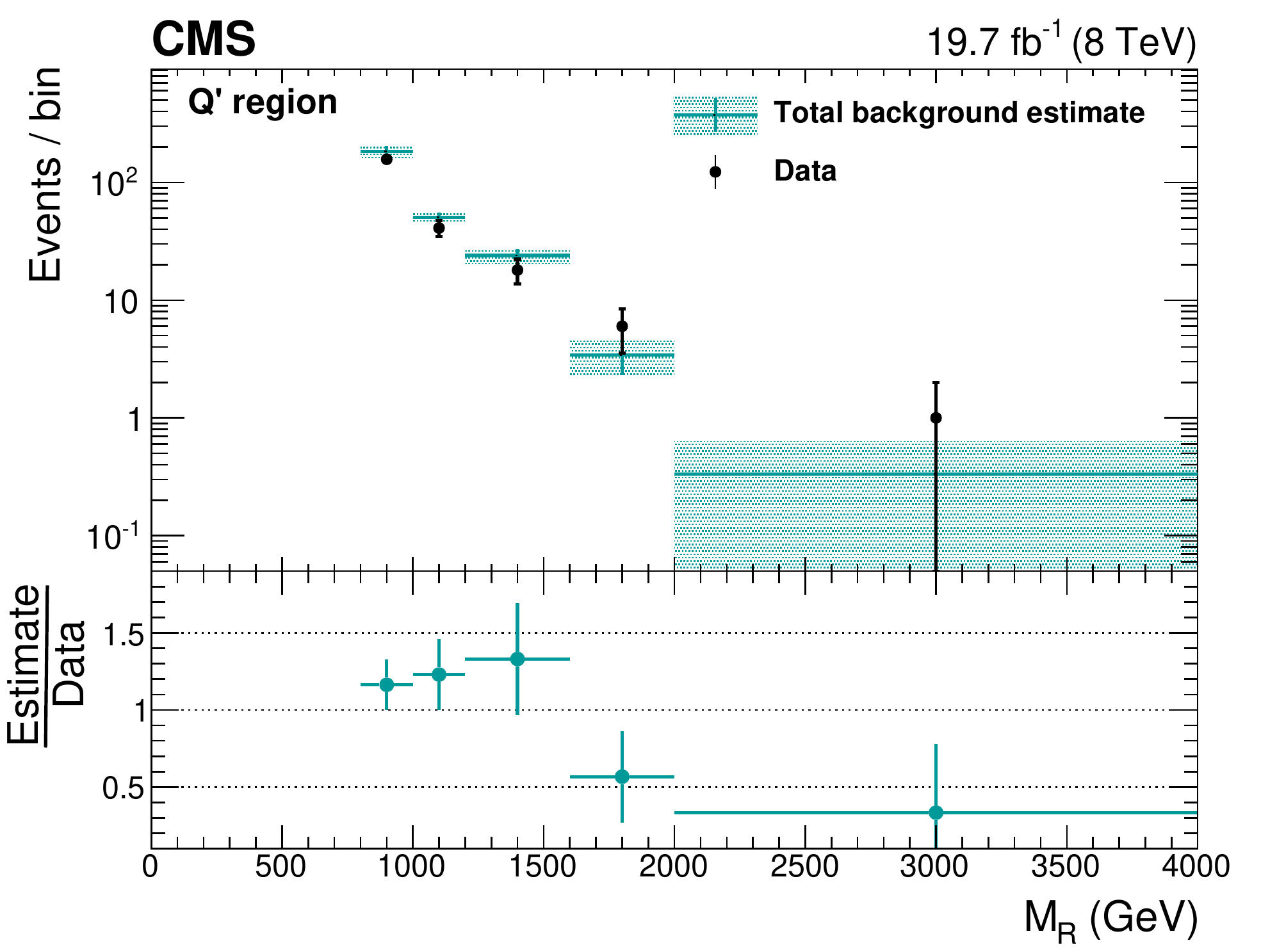}
\includegraphics[width=0.45\textwidth]{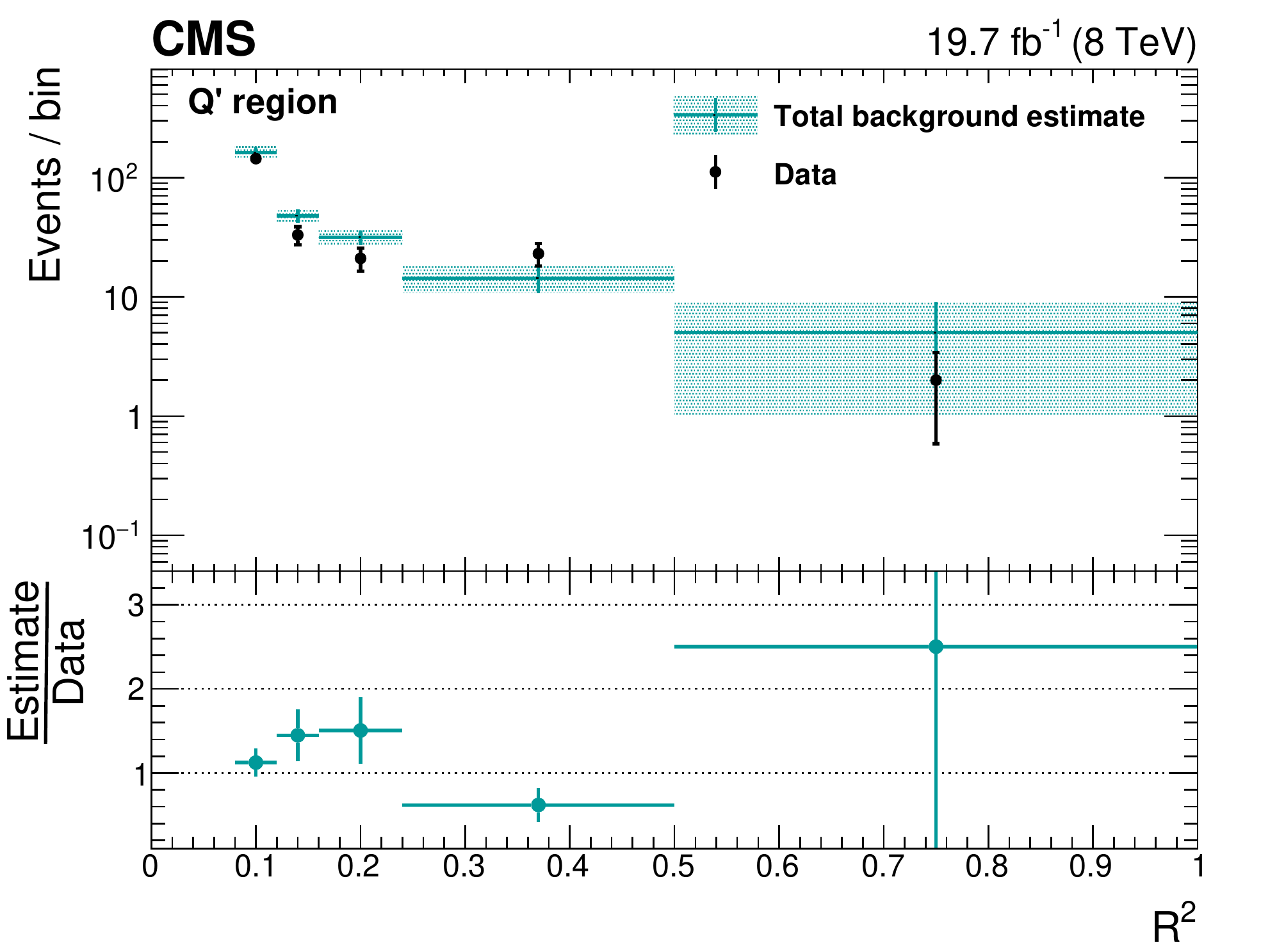}
\caption{One-dimensional projection of $\MR$ (left panel) and $R^2$ (right panel) for the cross-check predicting the background in region $Q'$ defined by $\Delta\phi_\text{min} > 0.5$.  The uncertainties shown are statistical only. The horizontal error bars indicate the bin width. \label{fig:Shape_syst_1D_project_QCD}}
\end{figure*}

In the second cross-check, we use the $Q$ region to estimate the background in a signal-like $Q$ region, denoted by $Q^\prime$, for which $\Delta\phi_\text{min} > 0.5$, from the relationship
\begin{equation}
  \widehat{N}^{Q^\prime} = N_\text{obs}^Q \frac{N_\mathrm{MC}^{Q^\prime}}{N_\mathrm{MC}^Q}.
\end{equation}
Here, $N_\mathrm{MC}$ includes all contributing background processes, and $N_\text{obs}^Q$ is the
observed count in the $Q$ region.
This test assesses the degree to which the simulated distribution of $\Delta\phi_\text{min}$ as well as its extrapolation from the $Q$ region to the $S$ region are reliable.
As observed from Table~\ref{tab:BG_comp_percent}, the multijet process is only a small contribution in the $Q'$ region. Therefore, this cross-check assesses how well the reduction of the multijet process, via the $\Delta\phi_\text{min}>0.5$ requirement, is modeled.
The comparison between prediction and observation can be made from data shown in Fig.~\ref{fig:Shape_syst_1D_project_QCD}.
 The level of discrepancy between the prediction
and the observation in this cross-check is incorporated as a systematic uncertainty of 42\% in the
global scale factor for the multijet component, as described in Section~\ref{sec:likelihood}.

\section{The \texorpdfstring{\PW}{W} boson tagging scale factors \label{sec:Wtag_SF}}

The $\PW$ boson tagger used in this analysis is the same as that defined and used in previous CMS analyses~\cite{EXO-12-024,Khachatryan:2014vla}.  Since the $\PW$ boson tagging efficiency does not depend significantly on the event topology, we use the same scale factor~\cite{EXO-12-024}
\begin{equation}
\textrm{SF}_{\textrm{Wtag}} = 0.86 \pm 0.07 ,
\end{equation}
as used in these previous analyses,
for correcting the modeling differences between \textsc{FullSim} and data for the $\PW$ boson tagging efficiency  and apply the scale factor to processes with genuine hadronically decaying $\PW$ bosons (mainly $\cPqt\cPaqt$ and signal) in the $S$ and $T$ regions.

On the other hand, the data/\textsc{FullSim} scale factors for the misidentification (mistag) efficiency for mass-tagged, antitagged, and tagged $\PW$ bosons are derived specifically for this analysis. The mistag efficiency is defined as the probability to tag, with one of the $\PW$ taggers, a jet not originating from the hadronic decay of a $\PW$ boson.
Scale factors are necessary to correct the mistag efficiencies for $\PW$ boson mass tagging and antitagging in the MC simulation of the $Q$ and $W$ control regions, respectively, whereas the mistag efficiency scale factor for $\PW$ boson tagging is used to correct simulated events with misidentified $\PW$ bosons, \eg multijet or $\PW({\to}\,\ell\PGn)$+jets events, in the $S$ and $T$ regions.
All three mistag efficiency scale factors are derived using the same multijet-enriched control region, defined as region $Q$ with the exception of all selections related to razor variables and $\PW$ tagging.
To obtain the mistag efficiencies $\epsilon_\mathrm{f}$ for $\PW$ boson tagging, mass tagging and antitagging, we use the leading CA8 jet in each event and measure the fraction of these jets passing the given tagger.
After obtaining $\epsilon_\mathrm{f}$ in both data and \textsc{FullSim}, we compute the scale factor,
\begin{equation}
\textrm{SF}(\pt) = \frac{\epsilon_\mathrm{f}^{\textrm{data}}(\pt)}{\epsilon_\mathrm{f}^{\textsc{FullSim}}(\pt)}.
\end{equation}
The scale factors for the $\PW$ boson tagging, mass tagging, and antitagging mistag efficiency vary between $1.0$--$1.2$, $1.1$--$1.4$, and $1.2$--$1.5$, respectively, depending on the CA8 jet \pt.
The uncertainties in the scale factor include the statistical uncertainty as well as the trigger efficiency and jet energy scale uncertainties, and vary between 2--7\% depending on the CA8 jet \pt.

Because the signal processes are simulated with \textsc{FastSim}, the resulting tagging efficiencies must be corrected for modeling differences between the programs \textsc{FastSim} and \textsc{FullSim}.
To compute the $\PW$ boson tagging efficiency \textsc{FullSim}/\textsc{FastSim} scale factor we use a sample of $\cPqt\cPaqt$ events simulated with \textsc{FullSim} and \textsc{FastSim}.
We first determine the $\PW$ boson tagging efficiency for both samples, considering only events with exactly one hadronically decaying $\PW$ boson at the generator level for which the closest reconstructed CA8 jet lies within $\Delta R = 0.8$ of the $\PW$ boson.
Since we wish to select boosted $\PW$ bosons, and not boosted top quarks, we require that there be no (generator-level) $\cPqb$ quark from the top quark decay within the cone of the closest CA8 jet.
The $\PW$ boson tagging efficiency as a function of \pt for a given sample is then obtained by dividing the \pt distribution of the closest CA8 jets that also satisfy the tagging condition ($70 < m_\text{jet} < 100\GeV$ and $\tau_{2}/\tau_{1} < 0.5$) by the \pt distribution of all of the closest CA8 jets.
To determine the \textsc{FullSim}/\textsc{FastSim} scale factor for the $\PW$ boson tagging efficiency, we divide the efficiencies $\epsilon$ obtained from the \textsc{FullSim} and \textsc{FastSim} samples, $\textrm{SF}_{\textrm{Full/Fast}}(\pt) = \epsilon^{\textsc{FullSim}}(\pt)/\epsilon^{\textsc{FastSim}}(\pt)$.
This scale factor is applied to all signal samples and varies between 0.89--0.95, depending on the \pt of the given CA8 jet, with an uncertainty of less than 3\%.

\section{Statistical analysis \label{sec:likelihood}}

The statistical analysis of the observations in the signal region is based on a likelihood function, $L(\sigma)$, given by
\begin{equation}
\ifthenelse{\boolean{cms@external}}
{
\begin{split}
L(\sigma)  = \ \ \ & \int \rd\mathcal{L} \int \rd\vec{\theta}_1 \cdots \int \rd\vec{\theta}_M \\
 & \left[ \prod_{i=1}^M p(N^S_i | \sigma, \mathcal{L}, \vec{\theta}_i)  \right] \,
\pi(\vec{\theta}_1,\cdots,\vec{\theta}_M) \, \pi(\mathcal{L}) ,
\end{split}
}
{
L(\sigma)  = \int \rd\mathcal{L} \int \rd\vec{\theta}_1 \cdots \int \rd\vec{\theta}_M
 \left[ \prod_{i=1}^M p(N^S_i | \sigma, \mathcal{L}, \vec{\theta}_i)  \right] \,
\pi(\vec{\theta}_1,\cdots,\vec{\theta}_M) \, \pi(\mathcal{L}) ,
}
\label{eq:marginal}
\end{equation}
where $\sigma$ is the total signal cross section, $M = 25$ is the number of bins in the $(M_{\textrm{R}},R^2)$ plane,
$N^S_i$ is the observed count in bin $i$ of the signal region,
and the bin-by-bin parameters  $\epsilon$,  $b^S_\text{multijet}$, $b^S_\mathrm{TTJ}$,
$b^S_{\PW(\to\ell\PGn)}$, and $b^S_\text{other}$ are denoted collectively by
$\vec{\theta}$.
The parameter $\epsilon$ represents the $M$
signal efficiencies (including acceptance) for a given signal model, while
the bin-by-bin background parameters for a given background process in the $S$ region are denoted by $b^{S}_\text{process}$.
The function $\pi(\mathcal{L})$ is the integrated luminosity prior and $\pi(\vec{\theta}_1,\cdots,\vec{\theta}_M)$ is an evidence-based prior
constructed from observations in the control regions and the four global scale factors $\kappa^{A/B}_\text{process} = \sum_i b^A_{\text{process}, \mathrm{MC}, i} / \sum_i b^B_{\text{process}, \mathrm{MC}, i}$, where the sum is over all bins of the simulated data; $A$ and $B$ denote any of the $S$, $Q$, $T$, or $W$ regions.

\begin{figure}[p]
\centering
\ifthenelse{\boolean{cms@external}}{
\includegraphics[width=0.48\textwidth]{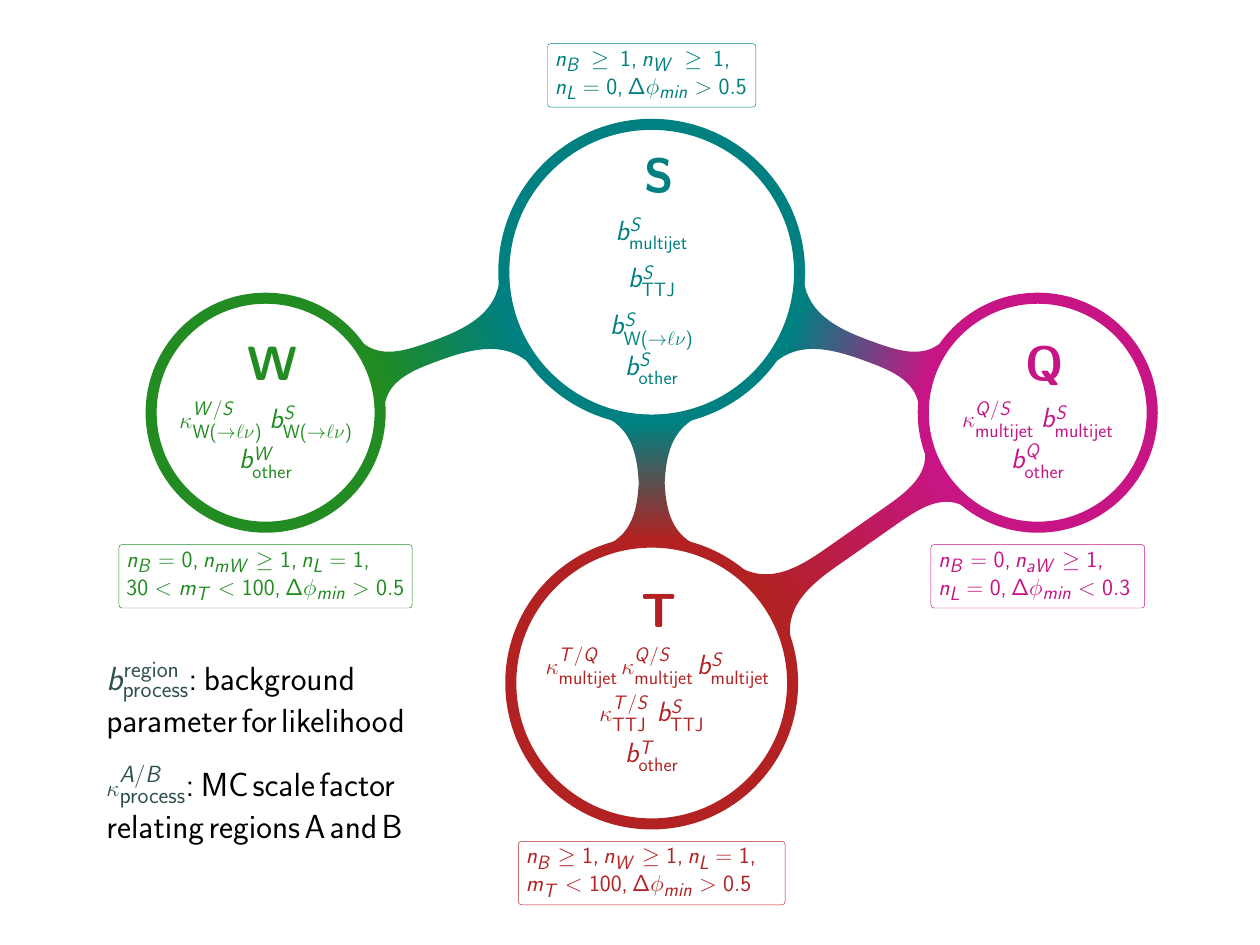}
}{
\includegraphics[width=0.75\textwidth]{Figure_009}
}
\caption{Graphical representation of the analysis method.
The circles represent the signal ($S$) and control ($Q,T,W$) regions, with their definition summarized in the associated boxes.
Listed inside each circle are the likelihood parameters relevant to that region: the bin-by-bin background parameters $b^\text{region}_\text{process}$ for the given region and background process, as well as the global scale factors $\kappa^{A/B}_\text{process} = \sum_i b^A_{\text{process}, \mathrm{MC}, i} / \sum_i b^B_{\text{process}, \mathrm{MC}, i}$, where the sum is over all bins of the simulated data.
A connection between two regions indicates that one or more parameters are shared.
The total expected background, per the $(\MR,R^2)$ bin, is the sum of the terms shown for each region.
Furthermore, associated with each bin of each region is an observed count, $N^\text{region}$, a simulated count, $N^\text{region}_{\text{process}, \mathrm{MC}}$, and
a count $N^\text{region}_{\text{other}, \mathrm{MC}}$ equal to the sum of the smaller backgrounds,
$\cPZ/\cPgg^*{\to}\,\ell\bar{\ell}+$jets, diboson, triboson, and $\cPqt\cPaqt V$,
with an associated parameter in the likelihood $b^\text{region}_\text{other}$.
\label{fig:BoostWorkflow}}
\end{figure}

The association of the global scale factors with the control regions is shown in
Fig.~\ref{fig:BoostWorkflow}, which also shows which control regions
provide constraints on the background parameters, $b^S_\text{  process}$. Although we use the same global scale factors in each
bin,  shape uncertainties in the simulated distributions are accounted
for by allowing the uncertainty in the scale factors to be bin dependent. The
25 signal bins in the $(M_{\textrm{R}},R^2)$ plane are divided into three sets for which different uncertainties are applied: the four bins nearest the origin
(set 1), the five surrounding bins (set 2), and the remaining bins
(set 3).
The likelihood per bin is taken to be
$p(N^S | \sigma, \mathcal{L}, \vec{\theta}) = \textrm{Poisson}(N^S,  \epsilon \sigma \mathcal{L} + b^S_\text{multijet} + b^S_\mathrm{TTJ} + b^S_{\PW(\to\ell\PGn)} +  b^{S}_\text{other})$.

The integral in Eq.~(\ref{eq:marginal}) is approximated using MC integration by sampling
 the priors $\pi(\mathcal{L})$ and $\pi(\vec{\theta}_1,\cdots,\vec{\theta}_M)$ and averaging the
 multibin likelihood with respect to the sampled points $\{(\mathcal{L}, \vec{\theta}_1,\cdots,\vec{\theta}_M)\}$.
The priors for the expected integrated luminosity $\mathcal{L}$, signal efficiencies $\epsilon$, and
simulated background counts $b^\text{region}_{\text{process}, \mathrm{MC}}$ are modeled with
gamma function densities,
\begin{align}
\mathrm{Ga}(x, \gamma, \beta) &= \beta^{-1}(x/\beta)^{\gamma-1} \exp(-x / \beta) / \Gamma(\gamma),
\label{eq:gamma}
\end{align}
in which the mode is set to $c$
and the variance to $\delta c^2$,
where
 $c \pm \delta c$ denotes either the measured integrated luminosity
 or,
for a given bin of a
 given region and process, the simulated signal efficiency,
 or the simulated background count. From $c \pm \delta c$,  we
 calculate
the gamma density parameters,
 \begin{align}
 	\gamma &= \Bigl[(k + 2) + \sqrt{(k+2)^2 - 4}\Bigr]/2,\\
	\beta &= \Bigl[\sqrt{c^2 + 4\delta c^2} - c\Bigr]/2,
 \end{align}
 where $k = (c / \delta c)^2$.
  For empty bins, we set $\gamma = 1$ and the bin value is
 constrained to zero by
 setting the $\beta$ parameter to $10^{-4}$.

For the signal efficiencies and backgrounds,
the prior is modeled
hierarchically,
\begin{equation}
\ifthenelse{\boolean{cms@external}}
{
\begin{split}
  \pi(\vec{\theta}_1,\cdots,\vec{\theta}_M) =
   \int \rd\vec{c}_1 & \cdots  \int \rd\vec{c}_M \int \rd\vec{\phi} \\    \left[ \prod_{i=1}^M \pi(\vec{\theta}_i | \vec{c}_i ) \right ] & \,
  \pi(\vec{c}_1,\cdots,\vec{c}_M |
  \vec{\phi} ) \pi(\vec{\phi}),
\end{split}
}
{
  \pi(\vec{\theta}_1,\cdots,\vec{\theta}_M) =
   \int \rd\vec{c}_1 \cdots \int \rd\vec{c}_M \int \rd\vec{\phi} \, \left[ \prod_{i=1}^M \pi(\vec{\theta}_i | \vec{c}_i ) \right ] \,
  \pi(\vec{c}_1,\cdots,\vec{c}_M |
  \vec{\phi} ) \pi(\vec{\phi}),
}
\label{eq:prior}
\end{equation}
where $\vec{\phi}$ represents parameters that characterize the independent
sources
of systematic uncertainty, described in Section~\ref{sec:systematics}.
The integral in Eq.~(\ref{eq:prior}) is
evaluated
as follows: $\vec{\phi}$
values are sampled from
$\pi(\vec{\phi})$ following the procedure described in Section~\ref{sec:systematics},
then $\vec{c}_{i}$ values from $\pi(\vec{c}_1,\cdots,\vec{c}_M | \vec{\phi})$, then $\vec{\theta}_i$ values from
$\pi(\vec{\theta}_i | \vec{c}_i)$. The sampling from $\pi(\vec{\phi})$ and $\pi(\vec{\theta}_i|\vec{c}_i)$
is straightforward because the functional forms are known. However,
the sampling of $\vec{c}_i$ requires running the analysis multiple times,
yielding an ensemble of
histograms in the $(\MR, R^2)$ plane, which
is the output of the procedure described in
Section~\ref{sec:systematics}. Thereafter, the sampling, which yields
the points
$\{(\mathcal{L}, \vec{\theta}, \cdots, \vec{\theta}_M)\}$, proceeds as follows:
\begin{enumerate}
\item sample the integrated luminosity parameter;
\item sample the efficiency parameters, $\epsilon$, for every bin and
  every signal model;
\item sample the background parameters $b^\text{region}_\text{process,
    MC}$ for
every bin and every background;
\item scale $b^Q_{\text{multijet, MC}}$ by a random number sampled
  from a gamma density of unit
  mode and standard deviation 0.36 in order to induce
the 42\% uncertainty in the multijet global scale factor $\kappa^{Q/S}_\text{multijet}$
that accounts for deficiencies in
the modeling of multijet production, as derived from the second cross-check mentioned in Section~\ref{sec:selection};
\item compute the $\kappa$ parameters from the appropriate
background sums, for example, $\kappa^{Q/S}_\text{multijet} =  \sum_i  b^Q_{\text{multijet}, \mathrm{MC}, i} /
\sum b^S_{\text{multijet}, \mathrm{MC}, i}$;
\item scale each $\kappa$ value by a
 random number sampled from a gamma density with unit mode  and
standard deviation of either 0.5 or 1.0 for the bins in set 2
or set 3, respectively, to account for the larger uncertainties in the
tails of the simulated distributions; and
\item sample the
background parameters $b^S_\text{multijet}$, $b^S_\mathrm{TTJ}$,
and
$b^S_{\PW(\to\ell\PGn)}$,
from the
Poisson
models  of the control regions; for example, for region $Q$,
$\textrm{Poisson}(N^Q ,  \kappa^{Q / S} b^S_\text{multijet} + b^Q_\text{  other})$ is mapped to a posterior density in $b^S_\text{multijet}$ using a
flat prior in $b^S_\text{multijet}$, and $b^S_\text{multijet}$ is sampled from the posterior density.
\end{enumerate}

If no statistically significant signal is observed, we determine limits on the total signal cross section using the CLs criterion~\cite{Junk:1999kv,Read:2002hq,LHCCLs} and
the test statistic $t_\sigma = 2 \ln [ L(\hat{\sigma}) /  L(\sigma)]$ when
$0 \leq\hat{\sigma} \leq \sigma$, and $t_\sigma = 0$ when $\hat{\sigma} > \sigma$. Large
values of $t_\sigma$ indicate incompatibility between the
best fit hypothesis $\sigma^\prime  = \hat{\sigma}$ and the hypothesis
$\sigma^\prime  = \sigma$ being tested. Given the $p$ values
$p_0 = \mathrm{Pr}(t_\sigma > t_{\sigma, \text{obs}} | \sigma^\prime = 0)$  and
$p_\sigma = \mathrm{Pr}(t_\sigma > t_{\sigma, \text{obs}} | \sigma^\prime=\sigma)$, obtained
by simulation, a 95\% CLs upper
limit on the cross section is obtained by solving
$\mathrm{CLs}(\sigma) = p_\sigma / p_0 = 0.05$. The
quantity $t_{\sigma, \text{obs}}$ denotes the
observed values of the test statistic, one for each hypothesis $\sigma^\prime=\sigma$.

\section{Systematic uncertainties \label{sec:systematics}}
The input to the statistical analysis is an ensemble of histograms in the $(\MR, R^2)$ plane that  incorporate systematic uncertainties in the simulated signal and background samples.   The independent systematic effects, described below, are sampled
simultaneously. For each sampled systematic effect,
a Gaussian variate with zero mean and unit variance is used in the calculation
of the random shift due to the systematic effect
 for all the signal and background models.
Likewise, the same randomly sampled PDFs are used for all signal and background models. In this way, the statistical dependencies among all bins of the signal and background models are correctly, and automatically, modeled. The sampling of the systematic effects is repeated several hundred times.

In all cases, except for those associated with PDFs, the systematic uncertainties are in the scale factors (SF)
applied to the simulated samples to correct them for modeling deficiencies. We consider the systematic uncertainties in the following quantities:

\begin{itemize}

\item {\bf Jet energy scale:}   The uncertainties are dependent on jet \pt and $\eta$~\cite{Chatrchyan:2011ds}.

\item {\bf Parton distribution functions:} We use 100 randomly sampled sets of PDFs from NNPDF23\_lo\_as\_0130\_qed~\cite{nnpdf},  MSTW2008lo68cl~\cite{Martin:2009iq}, and CT10~\cite{Lai:2010vv}.  The samples for the latter two are generated using the program
{\sc hessian2replicas}, recently
released with {LHAPDF6}~\cite{LHAPDF6}. Given a sampled set $i$, for PDF set $K$ and the
PDF set $O$ with which the events were simulated, events are reweighted using the scale factors,
$\mathrm{SF}_{K, i} = w_{K, i} / w_{O}$,
where the weights $w$ are products of the event-by-event PDFs for the colliding partons.

\item {\bf Trigger efficiency: }  We take the uncertainty in each bin, as a function of $\HT$ and leading jet $\pt$, to be the maximum of the statistical uncertainty in the efficiency after the baseline selection and the difference between the efficiencies before and after the baseline selection.

\item {\bf $\cPqb$ tagging scale factors:} The $\cPqb$ tagging performance differs between data and simulation, and differs between \textsc{FullSim} and \textsc{FastSim}, which is used to model signal processes.  The simulated events are therefore corrected by applying jet flavor-, \pt-, and $\eta$-dependent data/\textsc{FullSim} and \textsc{FullSim}/\textsc{FastSim} scale factors on the $\cPqb$ tagging or mistagging efficiency. The uncertainties in these scale factors are also jet flavor, \pt, and $\eta$ dependent, and are of the order of a few percent~\cite{btag8TeV}.

\item {\bf $\PW$ tagging scale factors:} The $\PW$ boson tag efficiency, and the mistag efficiency for $\PW$ boson tagging, $\PW$ boson mass tagging, and $\PW$ boson antitagging differ between data and simulation, as well as between \textsc{FullSim} and \textsc{FastSim}.  Data/\textsc{FullSim} and \textsc{FullSim}/\textsc{FastSim} scale factors,  whose uncertainties are functions of jet $\pt$, are applied to the simulated samples.

\item {\bf Lepton identification:} For electrons, we use \pt- and $\eta$-dependent scale factors for the identification efficiency. The  uncertainties
are also \pt and $\eta$ dependent~\cite{Khachatryan:2015hwa}.  The scale factor for the muon identification efficiency equals one and the corresponding uncertainties are negligible~\cite{Chatrchyan:2012xi}.

\item {\bf Initial-state radiation:} Deficiencies in the modeling of ISR
are corrected by reweighting~\cite{Chatrchyan:2013xna} the signal samples using an event weight
that depends on the \pt of the recoiling system.  The associated systematic uncertainty is equal to the difference $1 - w_\mathrm{ISR}$, where $w_\mathrm{ISR}$ is the ISR event weight.

\item {\bf Top quark transverse momentum:} Differential top quark pair production cross section analyses have shown that the shape of the \pt spectrum of  top quarks in data is softer than predicted~\cite{toppt}. To account for this, we reweight events  based on the \pt of the generator level $\cPqt$ and $\cPaqt$ quarks in the $\cPqt\cPaqt$ simulation.
The uncertainty associated with this reweighting is taken to be equal to the full amount of the reweighting.

\item {\bf Pileup: } Simulated events are reweighted so that their vertex multiplicity distribution matches that observed in data. The minimum-bias cross section is varied by ${\pm} 5\%$, thereby changing the shape of the vertex multiplicity distribution and therefore the weights.

\item {\bf Multijet spectrum:} The cross-checks described in Section~\ref{sec:selection} showed that there is a 42\% uncertainty in the multijet scale factor $\kappa$ between the $S$ and $Q$ regions.  This uncertainty is incorporated by increasing the uncertainty in the $\kappa$
parameter, as described in Section~\ref{sec:likelihood}.

\item {\bf $\cPZ ({\to}\, \PGn \PAGn)+$jets prediction:} About 8\% of the background in the signal region is composed of $\cPZ({\to}\,\PGn\PAGn)$+jets events. Since we require the presence of at least one $\cPqb$-tagged jet, and given the known deficiency in modeling $\cPZ$ production in association with heavy flavor quarks~\cite{Chatrchyan:2014dha}, we include an extra systematic uncertainty in the $\cPZ({\to}\,\PGn\PAGn)$+jets contribution.  This uncertainty is estimated using a data control region enriched in $\cPZ({\to}\,\ell \bar{\ell})+$jets, required to have exactly two tight leptons with the same flavor ($\Pe$ or $\PGm$) and opposite charge, $60 < m_{\ell\bar{\ell}} < 120\GeV$, at least one $\cPqb$-tagged jet, and at least one $\PW$ mass-tagged jet.  We estimate the uncertainty by first computing bin-by-bin data/simulation ratios in this control region.  Then, we take the uncertainty in the ratio in each bin as the standard deviation of a Gaussian density, normalized to  the number of events in that bin.  Finally, the Gaussian
densities from all bins are superposed, and the uncertainty is taken to be the magnitude of the 68\% band around a ratio of unity.

\end{itemize}

As noted above, all systematic effects are varied simultaneously across $(\MR, R^2)$ bins. However,
to assess the effect of each systematic uncertainty individually, each one is varied by one standard deviation up and down.
The effect on the background count and signal efficiency in the signal region is shown in Table~\ref{tab:bgsigsys}.
The signal values are obtained from averaging over all mass points in the T1ttcc model ($\Delta m = 25\GeV$) plane.
The PDF systematic uncertainties are obtained by running over 100 different members from the three PDF sets and fitting a Gaussian function to the efficiency distribution.
The last line in the table corresponds to the full sampling of the systematic uncertainties. To obtain this value, we again fit a Gaussian function to the efficiency distribution obtained from the full systematic sampling including 500 variations.
Although the effects of some of these systematic uncertainties on the backgrounds are large, they do not influence our results greatly because only the ratios of simulated background counts enter the statistical analysis, not the absolute values.  Therefore, most of the systematic effects cancel.
The statistical precision on the number of events in the control regions is the leading uncertainty in the background prediction for the search bins at large $\MR$ or $R^2$.
The dominant systematic uncertainty in the signal efficiency arises from the PDFs.

{
\begin{table*}[tpb]
\centering
\topcaption{Summary of $\pm 1$ standard deviation systematic uncertainties for the average signal efficiency over all mass assumptions in the T1ttcc model ($\Delta m=25\GeV$), and for the total background count in the signal region, unless indicated otherwise, as determined from simulation.  \label{tab:bgsigsys}}
\newcolumntype{x}{D{x}{\,}{-1}}
\begin{scotch}{l x x}
Systematic effect & \multicolumn{1}{c}{Signal (\%)} & \multicolumn{1}{c}{Background (\%)} \\
\hline
Jet energy scale &  +2.2 x {-2.1}   &  +10.9 x {-5.2}\\
Trigger &  +1.1 x {-3.3} &  +3.4 x {-5.7}\\
$\cPqb$ tagging \textsc{FullSim} &  +2.1 x {-2.3}& +3.9 x {-4.0}\\
$\cPqb$ tagging \textsc{FastSim} &  +1.2x  {-1.3}& \multicolumn{1}{c}{\NA} \\
$\PW$ tag efficiency {\sc Fullsim} &  +9.0 x {-8.9}& +4.6 x {-4.6}\\
$\PW$ tag efficiency \textsc{FastSim} &  +2.2 x {-2.2}&\multicolumn{1}{c}{\NA}\\
$\PW$ tag mistag efficiency \textsc{FullSim} &\multicolumn{1}{c}{\NA}&  +1.4 x {-1.4} \\
$\PW$ antitag mistag efficiency \textsc{FullSim} ($Q$ region only) &\multicolumn{1}{c}{\NA}& +2.6 x {-2.6} \\
$\PW$ mass-tag mistag efficiency \textsc{FullSim} ($W$ region only) &\multicolumn{1}{c}{\NA}& +2.3 x {-2.3} \\
Electron identification ($T$ and $W$ region only) &\multicolumn{1}{c}{\NA}& +0.2 x {-0.2} \\
Pileup &  +0.5 x {-0.5} & +1.0 x {-1.1}\\
ISR &  +6.6 x {-6.6} &\multicolumn{1}{c}{\NA}\\
Top quark $\pt$ &\multicolumn{1}{c}{\NA}&  +20.5 x {-14.4} \\
$\cPZ(\to\PGn\PAGn)+$ heavy flavor  &\multicolumn{1}{c}{\NA}& +4.0 x {-4.0} \\
PDF & \multicolumn{1}{c}{$20.7$} &  \multicolumn{1}{c}{$10.7$} \\
\hline
All &  \multicolumn{1}{c}{$24.4$} &  \multicolumn{1}{c}{$22.1$} \\
\end{scotch}
\end{table*}
}

\section{Results and interpretation \label{sec:interpretation}}

\begin{figure*}[p]
\centering
\includegraphics[width=0.49\textwidth]{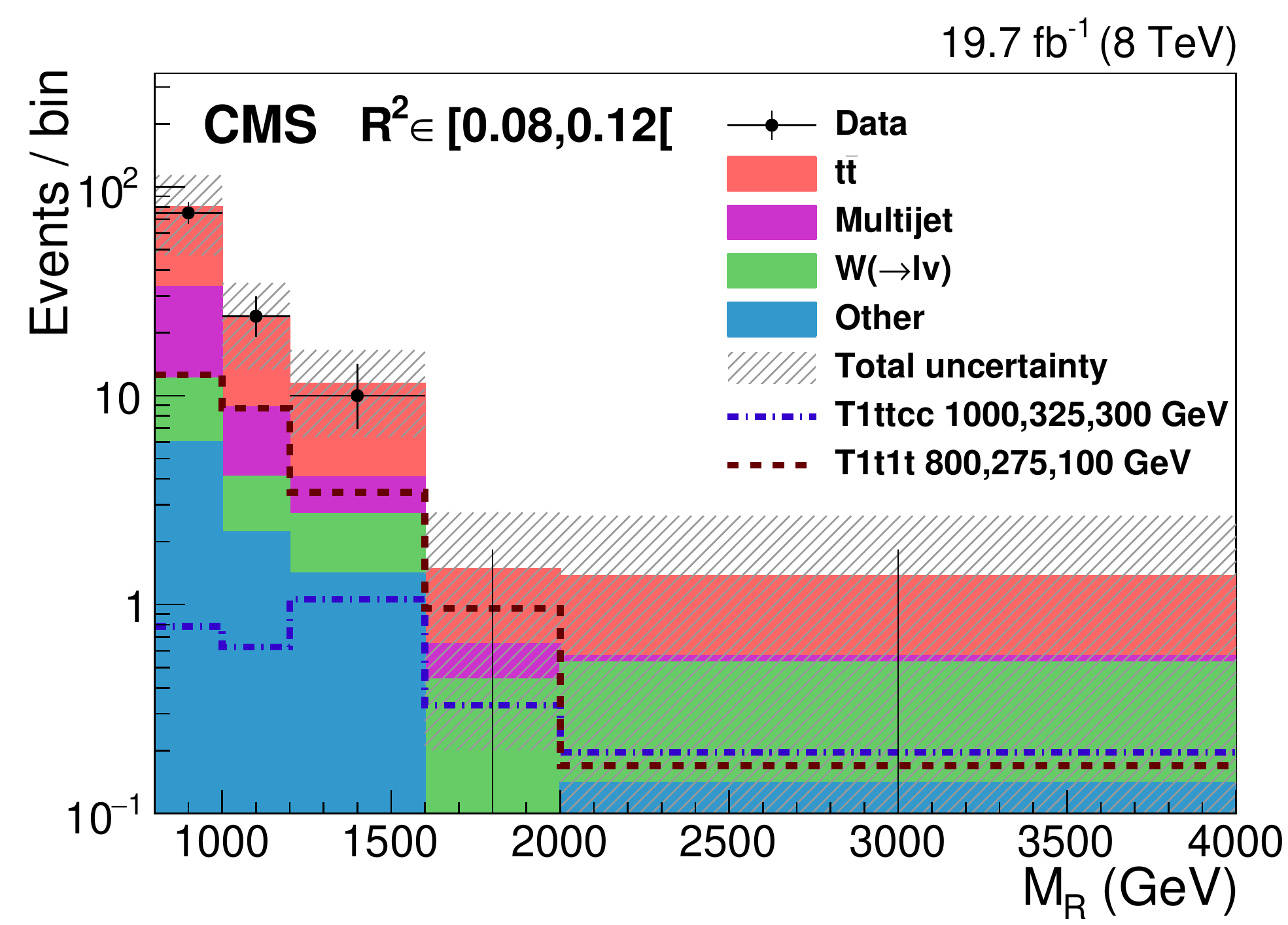}
\includegraphics[width=0.49\textwidth]{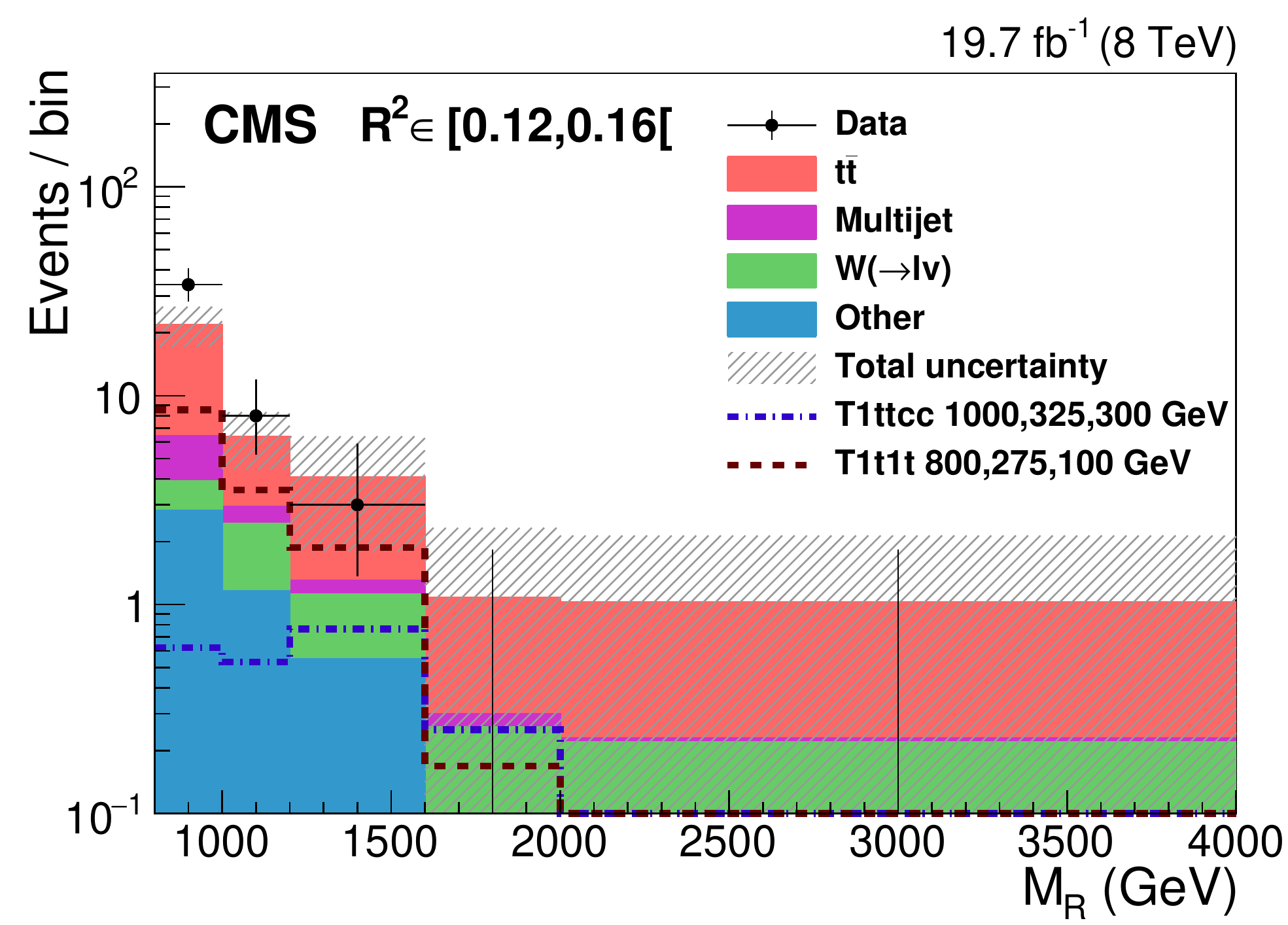}

\includegraphics[width=0.49\textwidth]{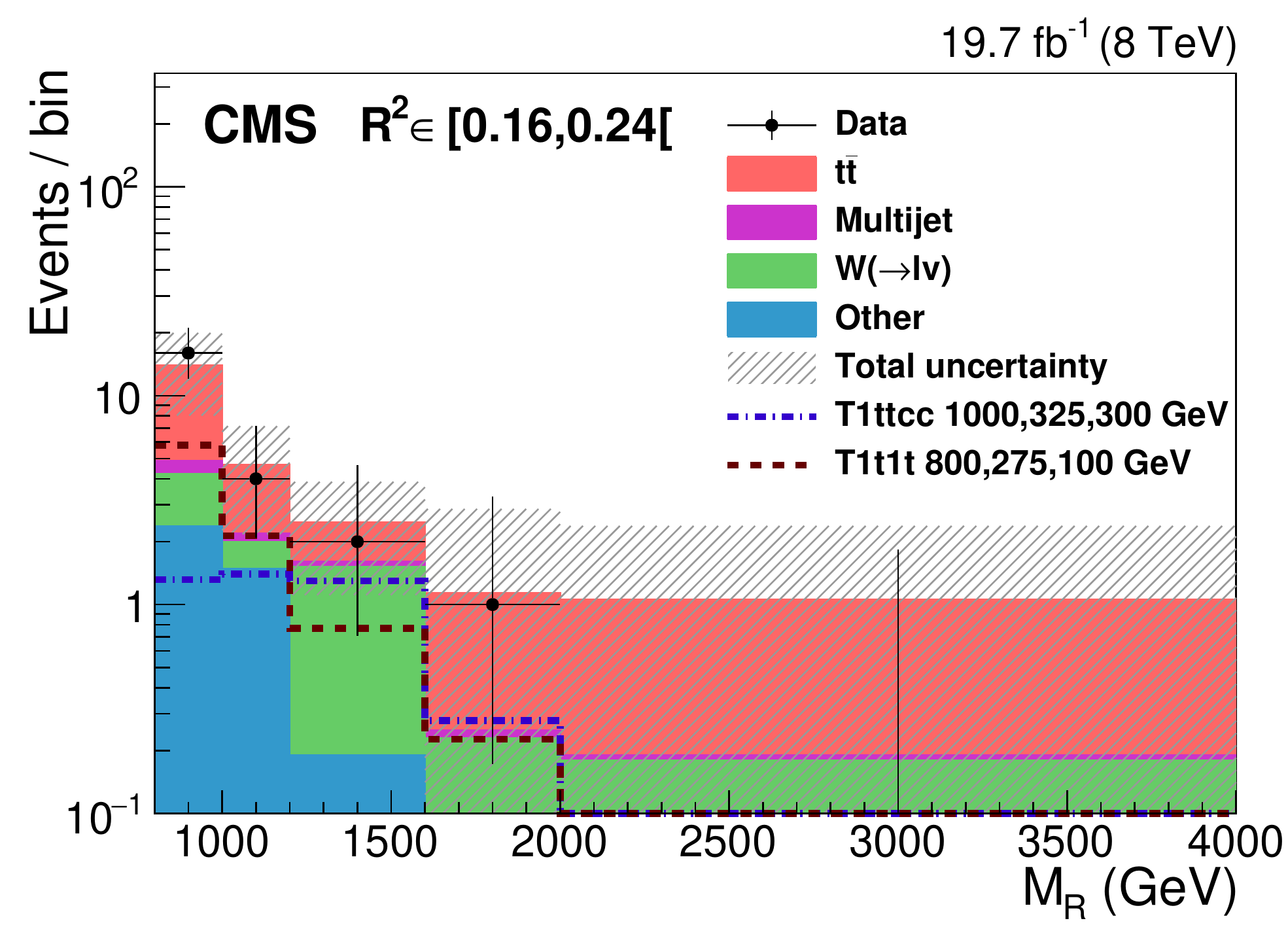}
\includegraphics[width=0.49\textwidth]{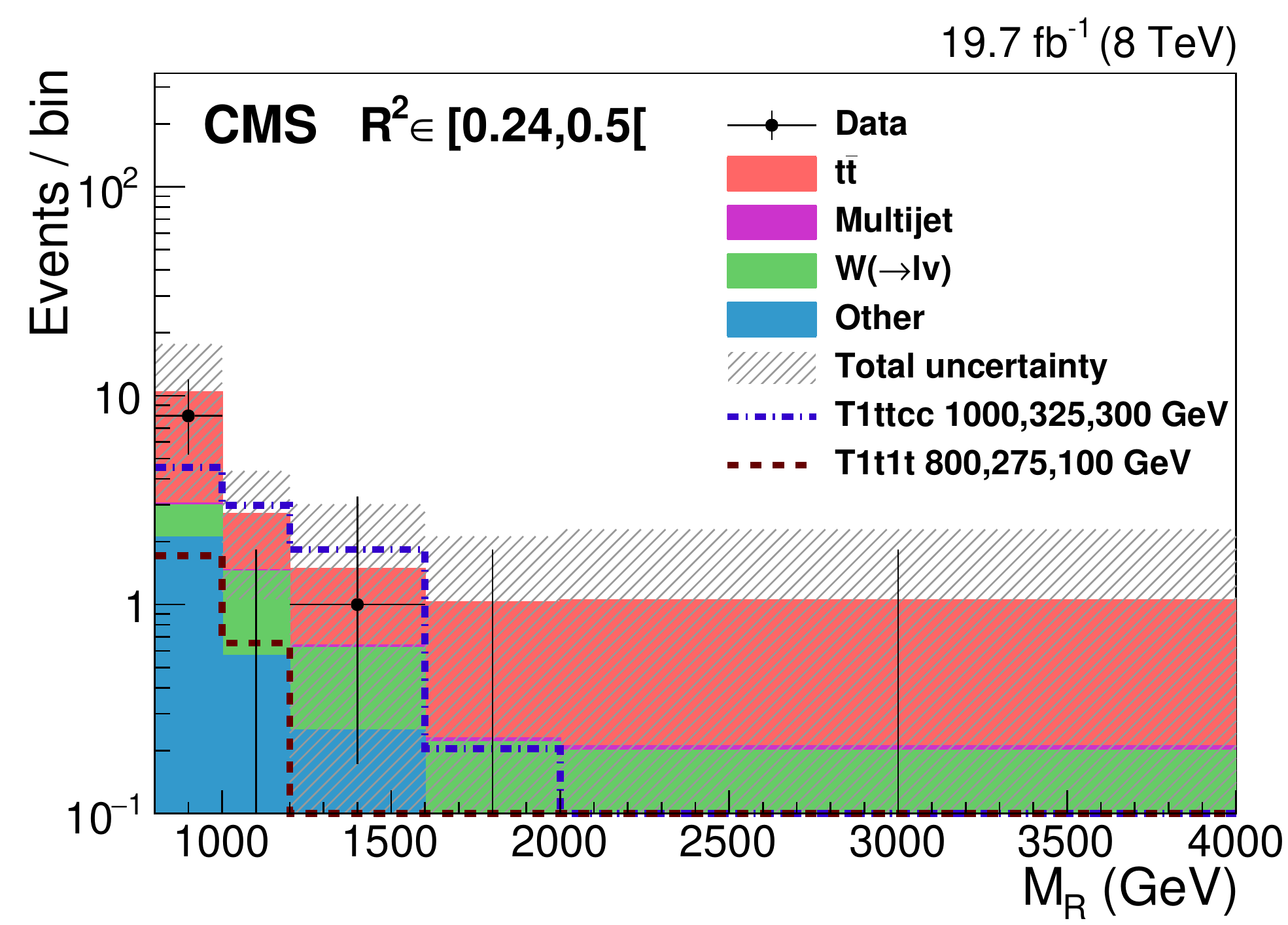}

\includegraphics[width=0.49\textwidth]{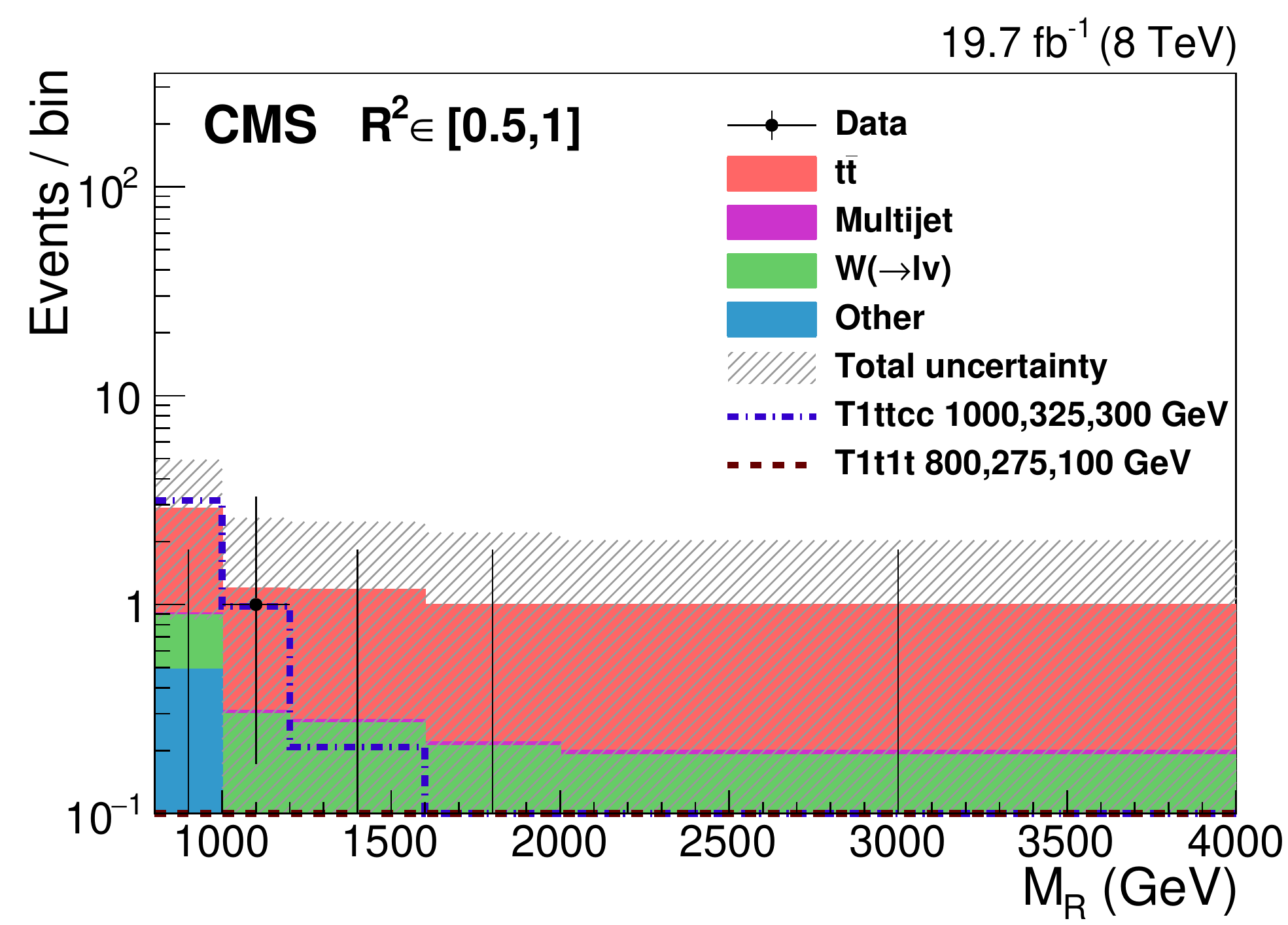}
\caption{Background predictions and observations. The results are shown in bins of $\MR$ for each $R^2$ bin.
The hatched band represents the total uncertainty in the background prediction.
Overlaid are two signal distributions corresponding to the T1ttcc model with $m_{\PSg} =1\TeV$, $m_{\PSQt} =325\GeV$, and $m_{\PSGczDo} =300\GeV$, and the T1t1t model with $m_{\PSg} =800\GeV$, $m_{\PSQt} =275\GeV$, and $m_{\PSGczDo} =100\GeV$.
\label{fig:results_prediction}}
\end{figure*}

Our background predictions for each bin in the
$(\MR,R^2)$ plane are presented in
Fig.~\ref{fig:results_prediction} and in
Table~\ref{tab:results_prediction}, which also lists the observed event yield in each bin.
The background predictions are presented as the mean and standard deviation as
determined from the background prior $\pi(\theta)$
described in Section~\ref{sec:likelihood}.
The observed event yields are found to be in agreement
with the predicted backgrounds from SM processes. Consequently, no evidence of a signal is observed.

\begin{table*}[p]
\centering
\topcaption{Event yields for the predicted backgrounds and for the data in each of the signal bins in $R^2$ and $\MR$.
The uncertainties in the predictions are the combined statistical and systematic uncertainties obtained using the sampling procedure described in the text.
\label{tab:results_prediction}}
\cmsTable{\small
\newcolumntype{y}{D{,}{\,\pm\,}{5,5}}
\newcolumntype{z}{D{,}{,\,}{7,7}}
\begin{scotch}{ c  z | y  y  y  y | y | c }
 \multicolumn{1}{c}{$R^2$} & \multicolumn{1}{c |}{$\MR$ (\GeVns)} &  \multicolumn{1}{c}{$\cPqt\cPaqt$} &  \multicolumn{1}{c}{Multijet} &  \multicolumn{1}{c}{$\PW({\to}\,\ell \PGn)$} &  \multicolumn{1}{c|}{Other} &  \multicolumn{1}{c|}{Total} & Observed\\
\hline \hline
\multirow{5}{*}{[0.08, 0.12[} & [800, 1000[ & 47.1 , 8.6 & 21.1 , 32.0 & 6.1 , 1.9 & 6.0 , 2.3 & 80.2 , 33.4 & 75 \\
 & [1000, 1200[ & 15.2 , 4.1 & 4.7 , 9.9 & 1.9 , 0.9 & 2.2 , 0.9 & 24.0 , 10.6 & 24 \\
 & [1200, 1600[ & 7.3 , 4.8 & 1.4 , 0.9 & 1.3 , 1.0 & 1.4 , 0.7 & 11.4 , 5.1 & 10 \\
 & [1600, 2000[ & 0.8 , 1.2 & 0.2 , 0.2 & 0.4 , 0.5 & 0.1 , 0.0 & 1.5 , 1.3 & 0 \\
 & [2000, 4000] & 0.8 , 1.1 & 0.0 , 0.1 & 0.4 , 0.6 & 0.1 , 0.1 & 1.4 , 1.3 & 0 \\
\hline
\multirow{5}{*}{[0.12, 0.16[} & [800, 1000[ & 15.5 , 4.2 & 2.5 , 1.2 & 1.1 , 0.8 & 2.8 , 1.2 & 21.9 , 4.8 & 34 \\
 & [1000, 1200[ & 3.4 , 1.8 & 0.5 , 0.3 & 1.3 , 0.6 & 1.2 , 0.7 & 6.4 , 2.0 & 8 \\
 & [1200, 1600[ & 2.8 , 2.3 & 0.2 , 0.1 & 0.6 , 0.5 & 0.6 , 0.4 & 4.1 , 2.3 & 3 \\
 & [1600, 2000[ & 0.8 , 1.2 & 0.0 , 0.1 & 0.2 , 0.3 & 0.1 , 0.0 & 1.1 , 1.2 & 0 \\
 & [2000, 4000] & 0.8 , 1.1 & 0.0 , 0.0 & 0.2 , 0.4 & 0.0 , 0.0 & 1.0 , 1.1 & 0 \\
\hline
\multirow{5}{*}{[0.16, 0.24[} & [800, 1000[ & 9.1 , 5.8 & 0.7 , 0.4 & 1.8 , 1.4 & 2.4 , 1.1 & 14.0 , 6.0 & 16 \\
 & [1000, 1200[ & 2.5 , 2.4 & 0.2 , 0.1 & 0.5 , 0.5 & 1.5 , 0.8 & 4.7 , 2.5 & 4 \\
 & [1200, 1600[ & 0.9 , 1.0 & 0.1 , 0.1 & 1.3 , 0.9 & 0.2 , 0.2 & 2.5 , 1.4 & 2 \\
 & [1600, 2000[ & 0.9 , 1.6 & 0.0 , 0.0 & 0.2 , 0.3 & 0.0 , 0.0 & 1.1 , 1.7 & 1 \\
 & [2000, 4000] & 0.9 , 1.3 & 0.0 , 0.0 & 0.2 , 0.3 & 0.0 , 0.0 & 1.1 , 1.3 & 0 \\
\hline
\multirow{5}{*}{[0.24, 0.5[} & [800, 1000[ & 7.4 , 7.0 & 0.1 , 0.1 & 0.9 , 1.2 & 2.1 , 1.0 & 10.4 , 7.2 & 8 \\
 & [1000, 1200[ & 1.3 , 1.4 & 0.0 , 0.0 & 0.9 , 1.0 & 0.6 , 0.3 & 2.7 , 1.6 & 0 \\
 & [1200, 1600[ & 0.8 , 1.4 & 0.0 , 0.0 & 0.4 , 0.6 & 0.2 , 0.2 & 1.5 , 1.5 & 1 \\
 & [1600, 2000[ & 0.8 , 1.1 & 0.0 , 0.0 & 0.2 , 0.2 & 0.1 , 0.0 & 1.0 , 1.1 & 0 \\
 & [2000, 4000] & 0.8 , 1.2 & 0.0 , 0.0 & 0.2 , 0.3 & 0.0 , 0.0 & 1.1 , 1.2 & 0 \\
\hline
\multirow{5}{*}{[0.5, 1]} & [800, 1000[ & 2.0 , 1.9 & 0.0 , 0.0 & 0.4 , 0.6 & 0.5 , 0.3 & 2.9 , 2.0 & 0 \\
 & [1000, 1200[ & 0.9 , 1.3 & 0.0 , 0.0 & 0.2 , 0.4 & 0.1 , 0.1 & 1.2 , 1.4 & 1 \\
 & [1200, 1600[ & 0.9 , 1.2 & 0.0 , 0.0 & 0.2 , 0.3 & 0.1 , 0.1 & 1.2 , 1.3 & 0 \\
 & [1600, 2000[ & 0.8 , 1.1 & 0.0 , 0.0 & 0.2 , 0.5 & 0.0 , 0.0 & 1.0 , 1.2 & 0 \\
 & [2000, 4000] & 0.8 , 1.0 & 0.0 , 0.0 & 0.2 , 0.3 & 0.0 , 0.0 & 1.0 , 1.0 & 0 \\
\end{scotch}
}
\end{table*}

We interpret our results in terms of the simplified model spectra T1ttcc and T1t1t, whose diagrams are shown in Fig.~\ref{fig:diagrams}.
These models each have three mass parameters: the gluino, top squark, and LSP masses. The
mass of the gluino is varied between 600 and 1300\GeV and that of the LSP between 1 and 500\GeV, while the mass difference between the top squark and the LSP, $\Delta m$, is fixed at 10, 25, or 80\GeV for the T1ttcc model, and at 175\GeV for the T1t1t model. In both models the gluino is assumed to decay 100\% of the time into a top squark and a top quark.

\begin{figure*}[p]
 \centering
 \includegraphics[width=0.49\textwidth]{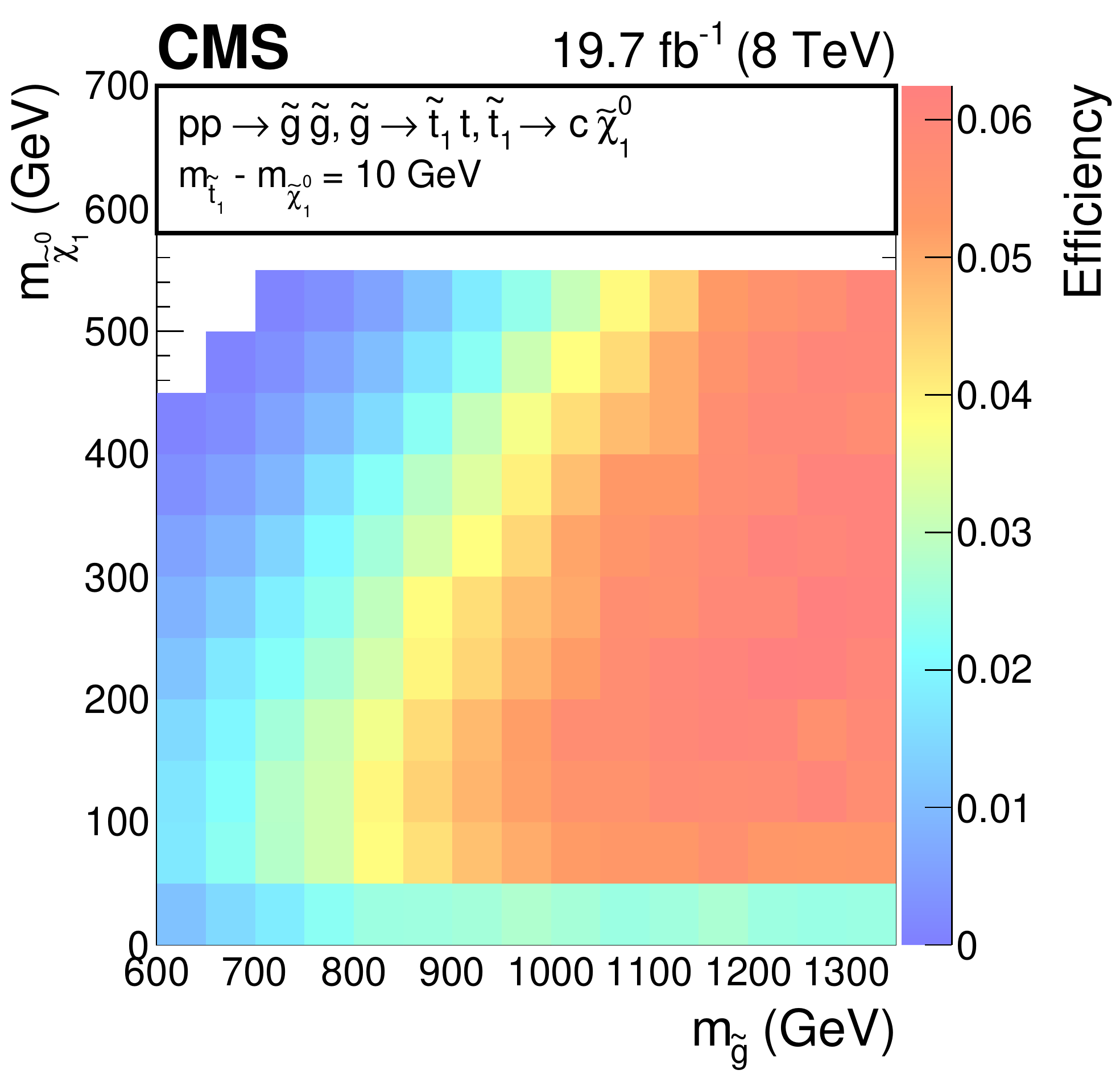} 
 \includegraphics[width=0.49\textwidth]{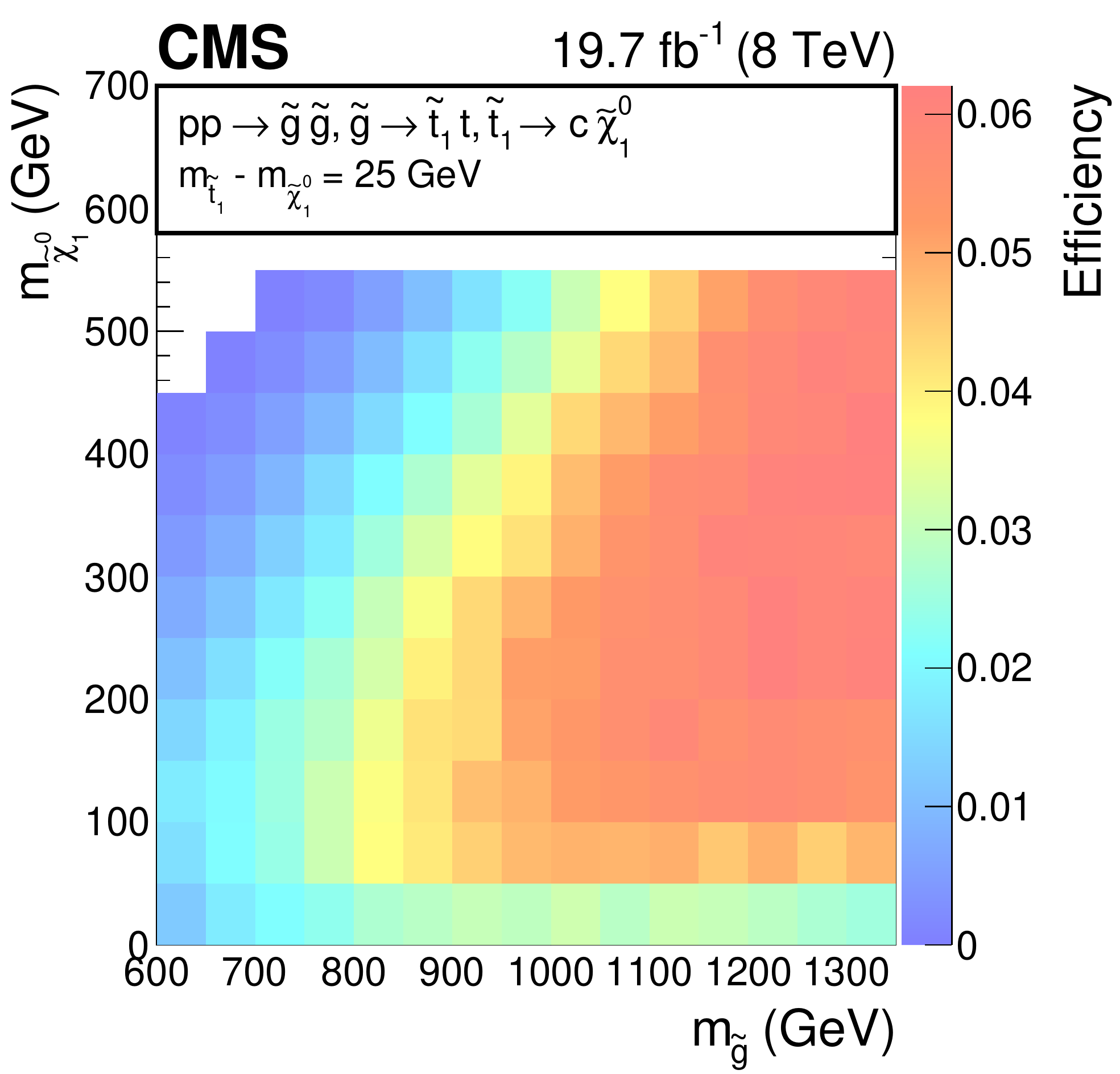}

 \includegraphics[width=0.49\textwidth]{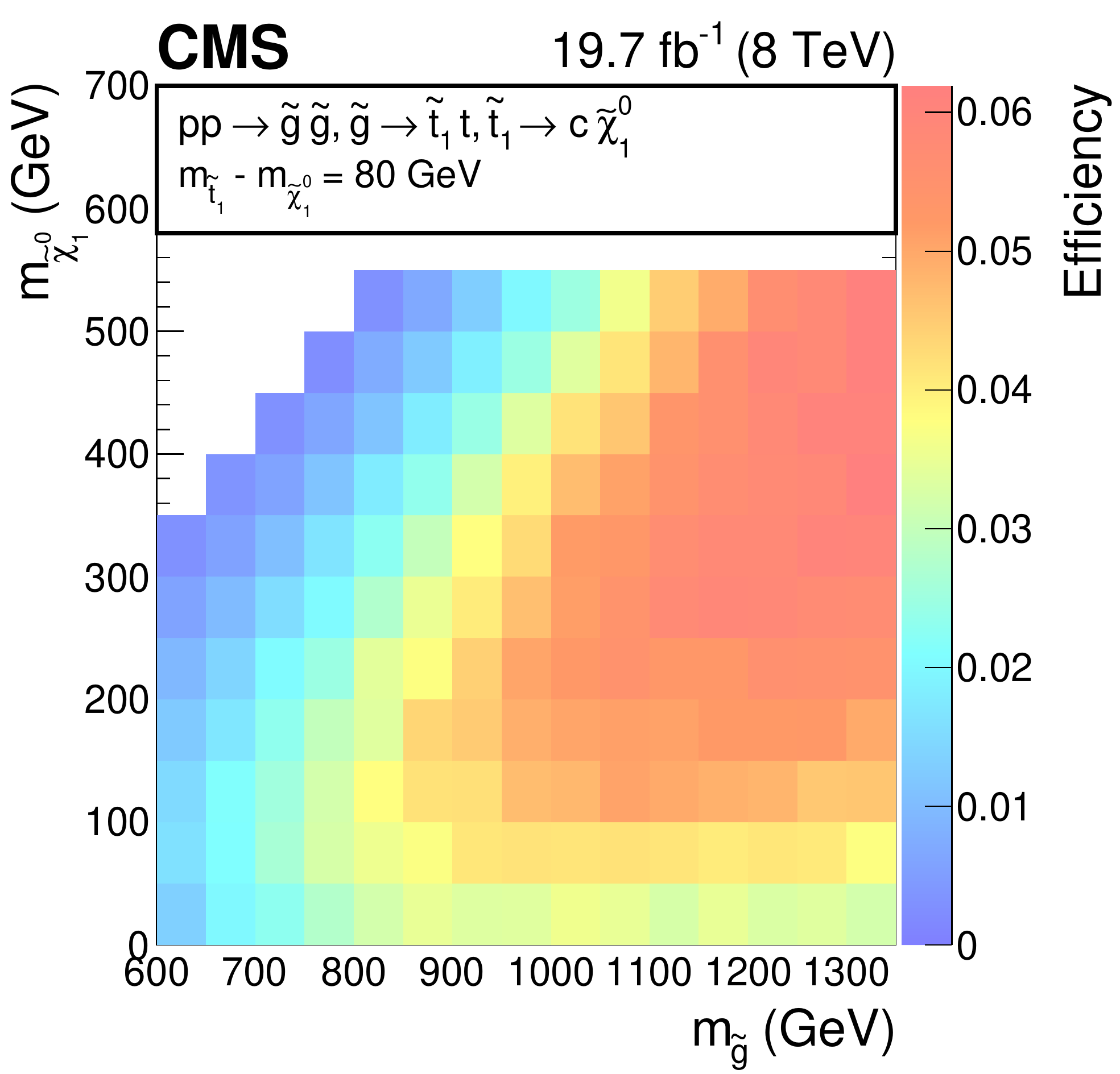} 
 \includegraphics[width=0.49\textwidth]{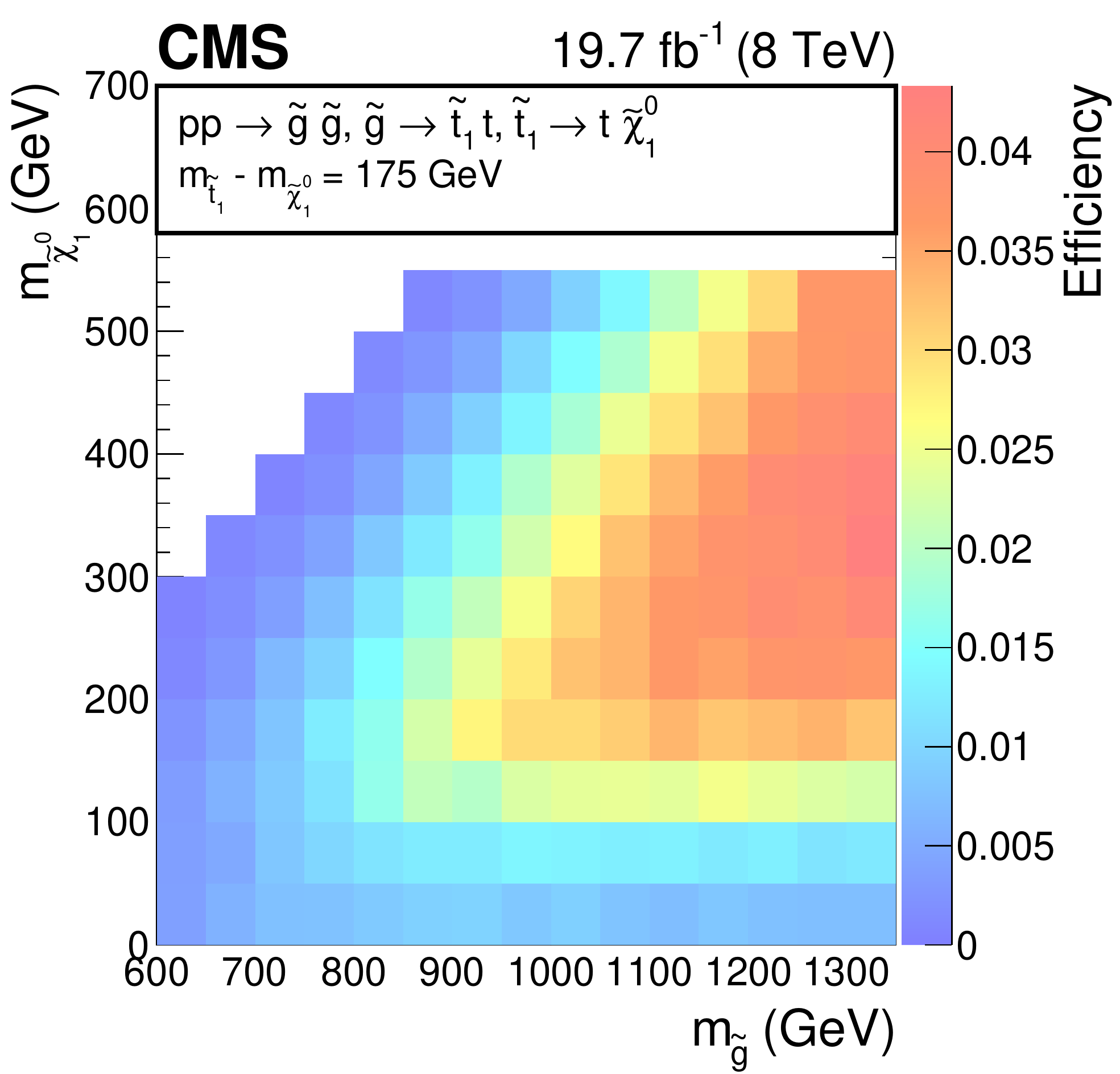}
 \caption{Signal efficiency for the T1ttcc and T1t1t simplified model spectra, as a function of the gluino and neutralino masses. Three mass splittings between top squark and LSP are considered for the T1ttcc model: 10, 25, and 80\GeV, shown in the top left, top right, and bottom left panels, respectively. The efficiency for the T1t1t model with  a mass splitting of 175\GeV is shown in the bottom right panel.
 \label{fig:eff_T1ttcc}}
\end{figure*}

To illustrate the expected signal sensitivity, we show in Fig.~\ref{fig:eff_T1ttcc} the signal efficiencies as a function of the gluino and neutralino masses, for the T1ttcc model, to which this analysis is particularly sensitive, and for the T1t1t model.
Efficiencies of up to 6\% in the most boosted regimes  are reached.
For the T1ttcc model a drop in efficiency is observed for the region of model parameter space with the lowest neutralino mass ($m_{\PSGczDo} = 1\GeV$), which can be explained by Lorentz boosts. For LSP masses higher than the mass of the charm quark, the LSP will assume most of the momentum.  For the bins with the lowest LSP mass, however, the LSP and the charm quark have about equal mass, so that after the boost they will share the momentum about equally.  This results in a softer \ETm spectrum and therefore a lower $R^2$ value, which reduces the efficiency substantially.

\begin{figure*}[p]
\centering
\includegraphics[width=0.49\textwidth]{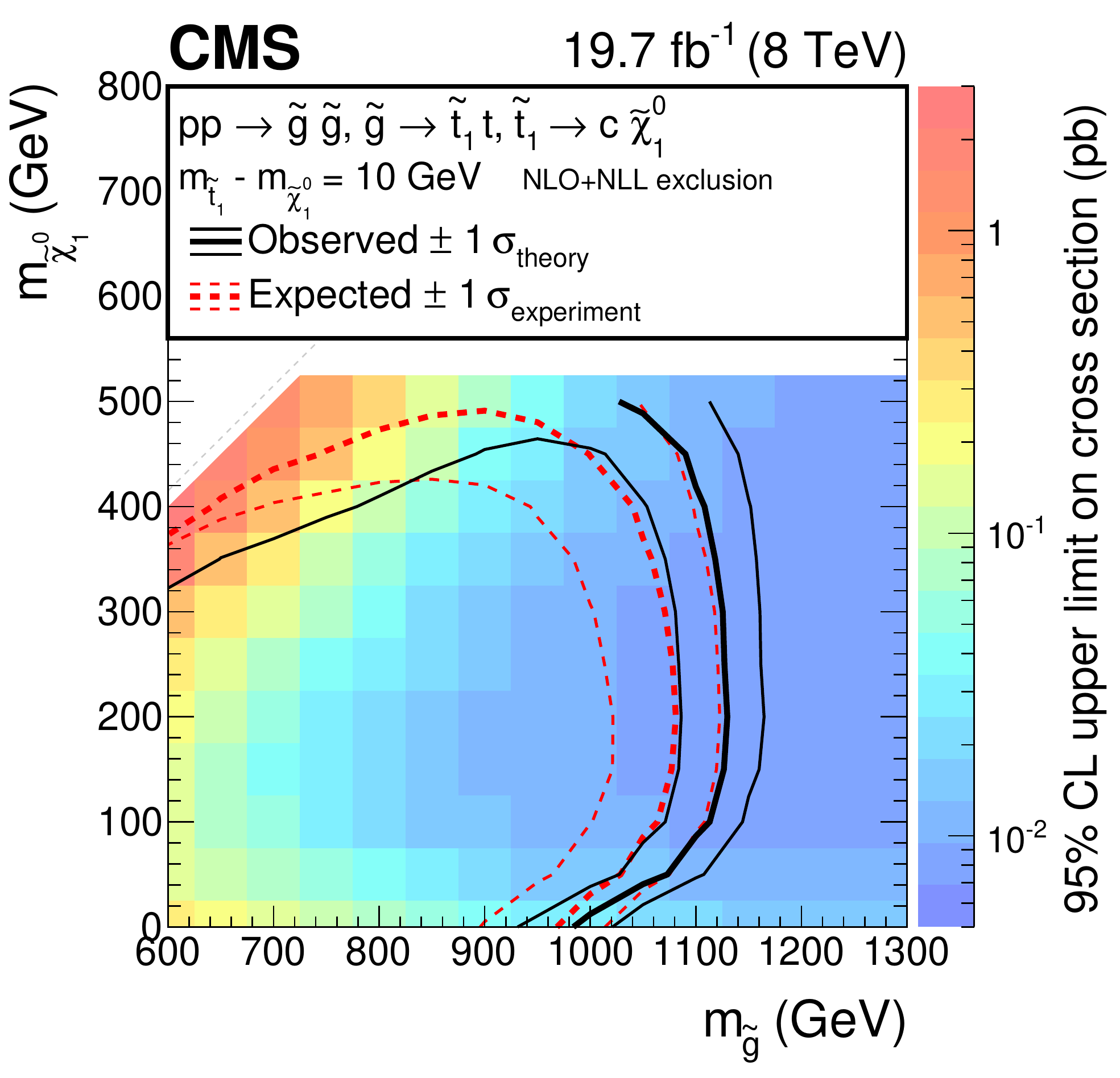}
\includegraphics[width=0.49\textwidth]{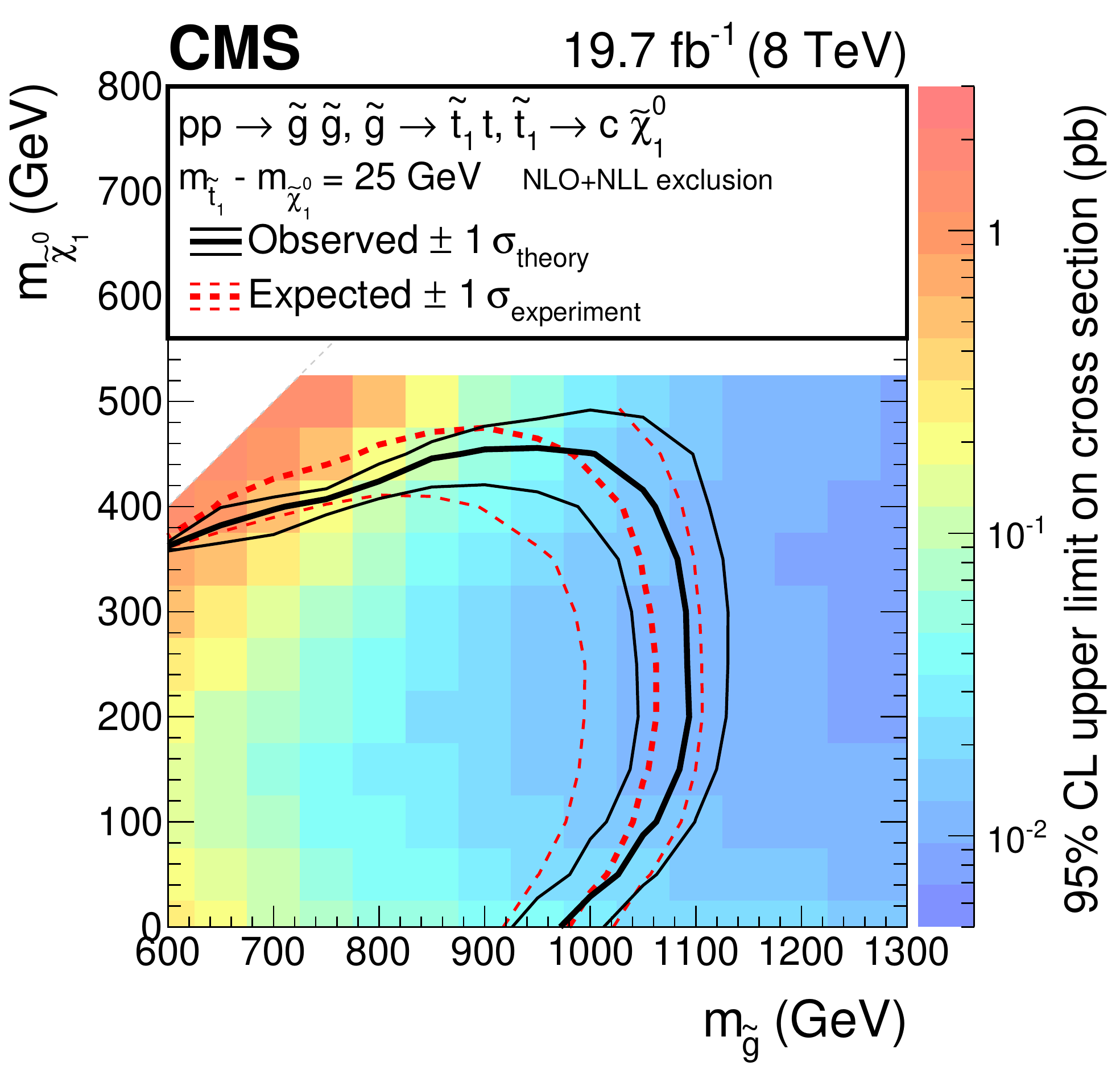}

\includegraphics[width=0.49\textwidth]{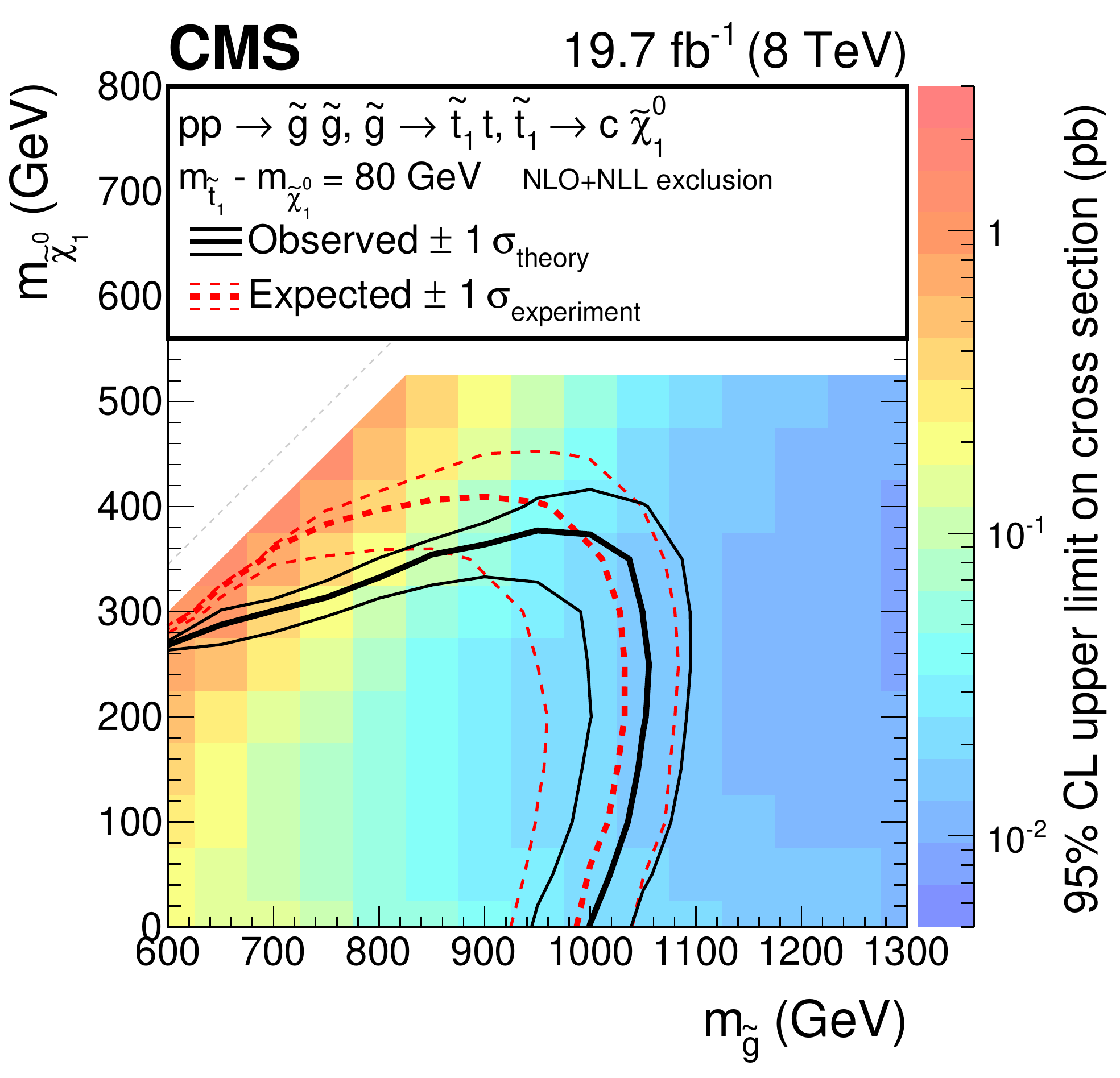}
\includegraphics[width=0.49\textwidth]{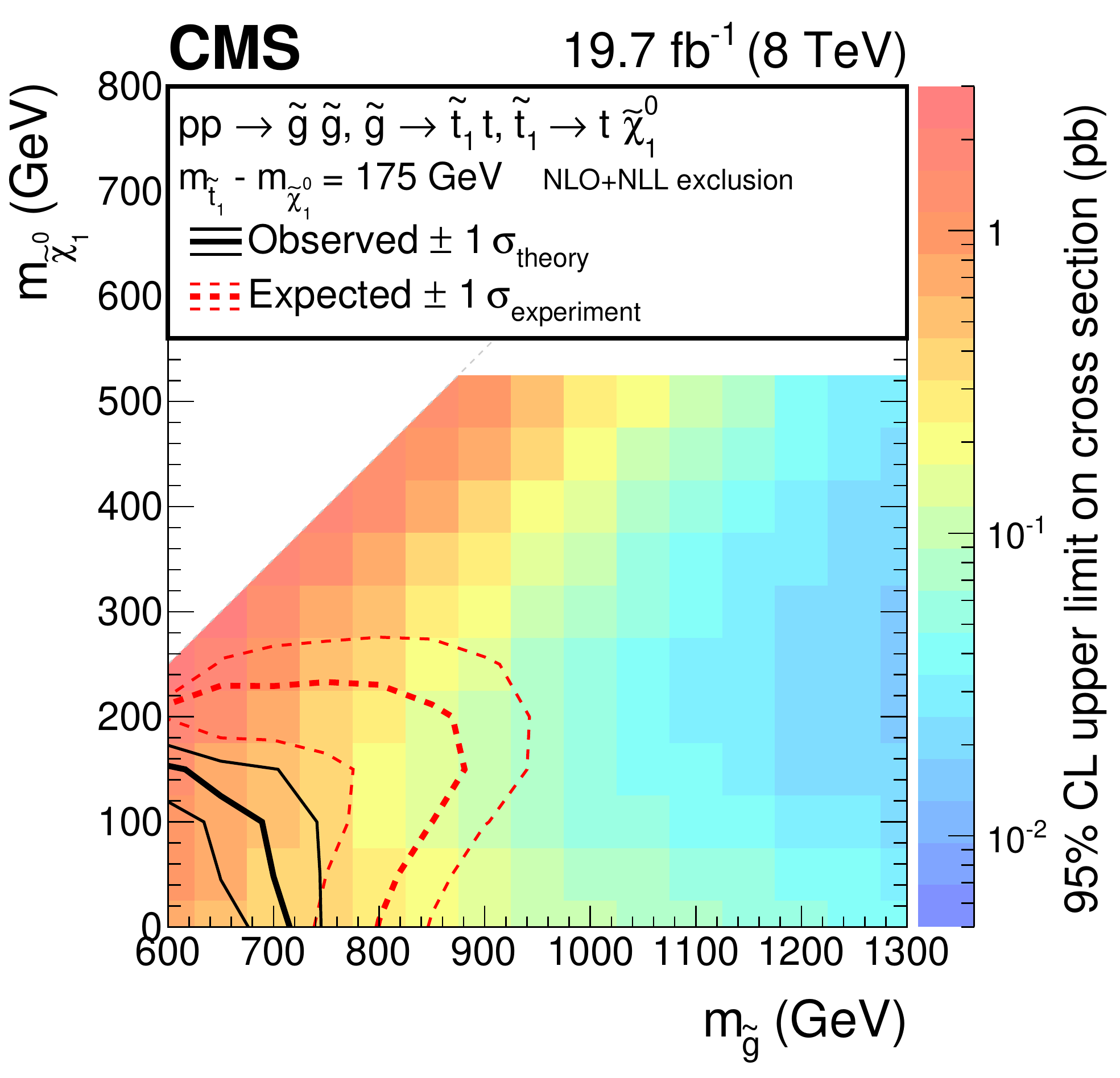}
\caption{Observed upper limit (CLs method, 95\% \CL) on the signal cross section as a function of the gluino
   and neutralino masses for the T1ttcc model with $\Delta m=10,$ 25, and 80\GeV (top left, top right, bottom left panels) and for the T1t1t model with $\Delta m = 175\GeV$ (bottom right panel).
Also shown are the contours corresponding to the observed and expected lower limits, including their uncertainties, on the gluino and neutralino masses.
\label{fig:limits}}
\end{figure*}

Figure~\ref{fig:limits} shows the observed 95\% confidence level (\CL) upper limit on the signal cross section as a function of the gluino
   and neutralino masses, obtained using the CLs method described briefly in
Section~\ref{sec:likelihood}, for the T1t1t model and for the T1ttcc model with $\Delta m=10, \,25, \,\textrm{and } 80$\GeV.
Additionally, the figure also shows contours corresponding to the observed and expected lower limits, including their uncertainties, on the gluino and neutralino masses.
This analysis has made significant inroads into the parameter space of the T1ttcc model.
Gluinos with mass up to about 1.1\TeV have been excluded for neutralinos with a mass less than about 400\GeV when the top squark decays to a charm quark and a neutralino and $\Delta m < 80\GeV$. This also means that top squarks with masses up to about 400\GeV have been excluded for small mass differences with the LSP, given the existence of a gluino with a mass less than about 1.1\TeV.
Similarly, for the T1t1t model, top squarks with a mass of up to about 300\GeV have been excluded for the scenarios with $\Delta m = 175\GeV$ and gluino mass less than 700\GeV.
The observed limit for this model is lower than the expected limit because of the small excess in the low $\MR$ bins for $0.12 \leq R^2 < 0.16$, which are among the most sensitive bins for the T1t1t model.

\section{Summary \label{sec:summary}}

We have presented a search for new physics in hadronic final states with at least one boosted $\PW$ boson and a $\cPqb$-tagged jet using data binned at high values of the razor kinematic variables, $\MR$ and $R^2$.  The analysis uses 19.7\fbinv of 8\TeV proton-proton collision data collected by the CMS experiment.
The SM backgrounds are estimated using control regions in data. Scale factors, derived from simulations, connect these control regions to the signal region.
The observations are found to be consistent with the SM expectation, as shown in Fig.~\ref{fig:results_prediction} and
Table~\ref{tab:results_prediction}.  The results, which are encapsulated in a binned likelihood, are interpreted in terms of supersymmetric models describing pair production of heavy gluinos decaying to boosted top quarks.  Limits are set on the gluino and neutralino masses using the CLs criterion on the gluino-neutralino mass plane, as shown in Fig.~\ref{fig:limits}.
Assuming that the gluino always decays into a top squark and a top quark, this analysis excludes gluino masses up to 1.1\TeV for top squarks with a mass of up to about 450\GeV that decay exclusively to a charm quark and a neutralino. In this scenario, the mass difference considered between the top squark and the neutralino is less than 80\GeV.
This analysis also excludes gluino masses of up to 700\GeV when the top squark decays solely to a top quark and a neutralino, and the mass difference between the top squark and the neutralino is around the top quark mass.

\section*{Acknowledgements}

\hyphenation{Bundes-ministerium Forschungs-gemeinschaft Forschungs-zentren} We congratulate our colleagues in the CERN accelerator departments for the excellent performance of the LHC and thank the technical and administrative staffs at CERN and at other CMS institutes for their contributions to the success of the CMS effort. In addition, we gratefully acknowledge the computing centres and personnel of the Worldwide LHC Computing Grid for delivering so effectively the computing infrastructure essential to our analyses. Finally, we acknowledge the enduring support for the construction and operation of the LHC and the CMS detector provided by the following funding agencies: the Austrian Federal Ministry of Science, Research and Economy and the Austrian Science Fund; the Belgian Fonds de la Recherche Scientifique, and Fonds voor Wetenschappelijk Onderzoek; the Brazilian Funding Agencies (CNPq, CAPES, FAPERJ, and FAPESP); the Bulgarian Ministry of Education and Science; CERN; the Chinese Academy of Sciences, Ministry of Science and Technology, and National Natural Science Foundation of China; the Colombian Funding Agency (COLCIENCIAS); the Croatian Ministry of Science, Education and Sport, and the Croatian Science Foundation; the Research Promotion Foundation, Cyprus; the Ministry of Education and Research, Estonian Research Council via IUT23-4 and IUT23-6 and European Regional Development Fund, Estonia; the Academy of Finland, Finnish Ministry of Education and Culture, and Helsinki Institute of Physics; the Institut National de Physique Nucl\'eaire et de Physique des Particules~/~CNRS, and Commissariat \`a l'\'Energie Atomique et aux \'Energies Alternatives~/~CEA, France; the Bundesministerium f\"ur Bildung und Forschung, Deutsche Forschungsgemeinschaft, and Helmholtz-Gemeinschaft Deutscher Forschungszentren, Germany; the General Secretariat for Research and Technology, Greece; the National Scientific Research Foundation, and National Innovation Office, Hungary; the Department of Atomic Energy and the Department of Science and Technology, India; the Institute for Studies in Theoretical Physics and Mathematics, Iran; the Science Foundation, Ireland; the Istituto Nazionale di Fisica Nucleare, Italy; the Ministry of Science, ICT and Future Planning, and National Research Foundation (NRF), Republic of Korea; the Lithuanian Academy of Sciences; the Ministry of Education, and University of Malaya (Malaysia); the Mexican Funding Agencies (CINVESTAV, CONACYT, SEP, and UASLP-FAI); the Ministry of Business, Innovation and Employment, New Zealand; the Pakistan Atomic Energy Commission; the Ministry of Science and Higher Education and the National Science Centre, Poland; the Funda\c{c}\~ao para a Ci\^encia e a Tecnologia, Portugal; JINR, Dubna; the Ministry of Education and Science of the Russian Federation, the Federal Agency of Atomic Energy of the Russian Federation, Russian Academy of Sciences, and the Russian Foundation for Basic Research; the Ministry of Education, Science and Technological Development of Serbia; the Secretar\'{\i}a de Estado de Investigaci\'on, Desarrollo e Innovaci\'on and Programa Consolider-Ingenio 2010, Spain; the Swiss Funding Agencies (ETH Board, ETH Zurich, PSI, SNF, UniZH, Canton Zurich, and SER); the Ministry of Science and Technology, Taipei; the Thailand Center of Excellence in Physics, the Institute for the Promotion of Teaching Science and Technology of Thailand, Special Task Force for Activating Research and the National Science and Technology Development Agency of Thailand; the Scientific and Technical Research Council of Turkey, and Turkish Atomic Energy Authority; the National Academy of Sciences of Ukraine, and State Fund for Fundamental Researches, Ukraine; the Science and Technology Facilities Council, UK; the US Department of Energy, and the US National Science Foundation.

Individuals have received support from the Marie-Curie program and the European Research Council and EPLANET (European Union); the Leventis Foundation; the A. P. Sloan Foundation; the Alexander von Humboldt Foundation; the Belgian Federal Science Policy Office; the Fonds pour la Formation \`a la Recherche dans l'Industrie et dans l'Agriculture (FRIA-Belgium); the Agentschap voor Innovatie door Wetenschap en Technologie (IWT-Belgium); the Ministry of Education, Youth and Sports (MEYS) of the Czech Republic; the Council of Science and Industrial Research, India; the HOMING PLUS program of the Foundation for Polish Science, cofinanced from European Union, Regional Development Fund; the OPUS program of the National Science Center (Poland); the Compagnia di San Paolo (Torino); MIUR project 20108T4XTM (Italy); the Thalis and Aristeia programs cofinanced by EU-ESF and the Greek NSRF; the National Priorities Research Program by Qatar National Research Fund; the Rachadapisek Sompot Fund for Postdoctoral Fellowship, Chulalongkorn University (Thailand); the Chulalongkorn Academic into its 2nd Century Project Advancement Project (Thailand); and the Welch Foundation, contract C-1845.

\clearpage

\bibliography{auto_generated}

\cleardoublepage \appendix\section{The CMS Collaboration \label{app:collab}}\begin{sloppypar}\hyphenpenalty=5000\widowpenalty=500\clubpenalty=5000\input{SUS-14-007-authorlist.tex}\end{sloppypar}
\end{document}

%% file: SUS-14-007-authorlist.tex
\textbf{Yerevan Physics Institute,  Yerevan,  Armenia}\\*[0pt]
V.~Khachatryan, A.M.~Sirunyan, A.~Tumasyan
\vskip\cmsinstskip
\textbf{Institut f\"{u}r Hochenergiephysik der OeAW,  Wien,  Austria}\\*[0pt]
W.~Adam, E.~Asilar, T.~Bergauer, J.~Brandstetter, E.~Brondolin, M.~Dragicevic, J.~Er\"{o}, M.~Flechl, M.~Friedl, R.~Fr\"{u}hwirth\cmsAuthorMark{1}, V.M.~Ghete, C.~Hartl, N.~H\"{o}rmann, J.~Hrubec, M.~Jeitler\cmsAuthorMark{1}, V.~Kn\"{u}nz, A.~K\"{o}nig, M.~Krammer\cmsAuthorMark{1}, I.~Kr\"{a}tschmer, D.~Liko, T.~Matsushita, I.~Mikulec, D.~Rabady\cmsAuthorMark{2}, N.~Rad, B.~Rahbaran, H.~Rohringer, J.~Schieck\cmsAuthorMark{1}, R.~Sch\"{o}fbeck, J.~Strauss, W.~Treberer-Treberspurg, W.~Waltenberger, C.-E.~Wulz\cmsAuthorMark{1}
% ...
\vskip\cmsinstskip
% ...
1:~~Also at Vienna University of Technology, Vienna, Austria\\
2:~~Also at CERN, European Organization for Nuclear Research, Geneva, Switzerland\\
% ...
